\documentclass[12pt]{iopart}
\pdfoutput=1
\expandafter\let\csname equation*\endcsname\relax
\expandafter\let\csname endequation*\endcsname\relax
\usepackage{color}
\usepackage[breakwords]{truncate}
\usepackage[font={small}]{caption}
\usepackage{physics}
\usepackage{graphicx}
\usepackage{cite} 
\usepackage{subcaption}
\usepackage{etoolbox}
\usepackage[section]{placeins}
\usepackage[sectionbib]{chapterbib}
\usepackage{amsfonts}
\usepackage[citecolor=magenta,colorlinks=true]{hyperref}
\makeatletter
\newrobustcmd{\fixappendix}{%
  \patchcmd{\l@section}{1.5em}{7em}{}{}%
  \patchcmd{\l@subsection}{2.3em}{7em}{}{}%
}
\makeatother
\appto\appendix{
\addtocontents{toc}{\fixappendix}
\addtocontents{toc}{\protect\setcounter{tocdepth}{1}}}

\begin{document}
\newcommand{\ben}[1]{\textcolor{blue}{\textbf{#1}}}
\title{Athermal fluctuations in three dimensional disordered crystals}
\author{Roshan Maharana}
\address{Centre for Interdisciplinary Sciences, Tata Institute of Fundamental Research, Hyderabad, India}
\eads{\mailto{roshanm@tifrh.res.in}}
\begin{abstract}
We study jammed near-crystalline materials composed of frictionless spheres in three dimensions. We analyze the fluctuations in positions and forces produced by small polydispersity in particle sizes. We generalize a recently developed perturbation expansion about the crystalline ordered state to three dimensional systems. This allows us to exactly predict changes in positions and forces as a response to the changes in particle radii. We show that fluctuations in forces orthogonal to the lattice directions are highly constrained as compared to the fluctuations along lattice directions. Additionally, we analyze the correlations in the displacement fields produced by the microscopic disorder, which we show displays long ranged behaviour. 
\end{abstract}

\newpage
{\pagestyle{plain}
 \tableofcontents
\cleardoublepage}

\section{Introduction}

A large class of materials composed of macroscopic particles where thermal agitations play a minute role in dynamics can be classified as athermal~\cite{bi2015statistical,jaeger1996granular,van2009jamming,henkes2007entropy,o2002random,o2003jamming,wyart2005rigidity,goodrich2012finite,ramola2017scaling,cates1998jamming,torquato2010jammed,berthier2011theoretical,kapteijns2019fast,bi2011jamming}. Jammed particle packings~\cite{bi2015statistical,jaeger1996granular,van2009jamming,henkes2007entropy,o2002random,o2003jamming,wyart2005rigidity,goodrich2012finite,ramola2017scaling,cates1998jamming,torquato2010jammed}, granular systems, low temperature glassy materials~\cite{berthier2011theoretical,kapteijns2019fast} as well as densely packed tissues~\cite{bi2011jamming} are examples of athermal systems. Such systems are non-ergodic even displaying complete dynamical arrest. These states of the system being governed by the local constraints of mechanical equilibrium, can be primarily described by their energy minimized states~\cite{tong2019revealing,vanderwerf2020pressure,hexner2018two}. Additionally, such systems exhibit behaviours not explained by the fluctuation-dissipation theorem, as the thermal motion of the particles are absent~\cite{grigera1999observation,kubo1966fluctuation} in comparison to the particle scale. Despite lacking internal dynamics these systems respond to  external perturbations such as shear or global stress which leads to the motion of the system in phase space. 
In order to describe such systems, several ensemble based frameworks have been proposed that seemingly reproduce aspects of disordered athermal materials~\cite{bi2015statistical, jaeger1996granular}. The mechanical response of athermal materials is intimately related to the microscopic disorder present in the system. In this context, extensions of the techniques of statistical mechanics to explain the microscopic properties of such materials represents a fundamentally new challenge.

Various athermal materials exhibit characteristic properties as a consequence of disorder at the particulate scale.  Even in the absence of thermal fluctuations, disordered materials show large deviations from crystalline behavior in correlations as well as the interparticle forces~\cite{jaeger1996granular,majmudar2005contact,acharya2020athermal,acharya2021disorder}. Disorder may be introduced in a system in various ways~\cite{ostoja2002lattice,nie2017role}, such as incorporating heterogeneity in the bond network by adding  randomness in particle sizes~\cite{tong2015crystals,o2002random,charbonneau2019glassy}, and by an addition of a quenched activity to the particles~\cite{acharya2021disorder,schwarz2013physics}. Yet other methods include introducing topological disorder by randomly creating vacancies by removing particles or bonds~\cite{nie2017role}, resulting in a loss of periodicity of the crystal. Many techniques have been developed to study crystals with discrete translational invariance~\cite{debye1912theory,montroll1942frequency,chaikin1995principles}, whereas systems with disorder are not amenable to similar treatments. However, disordered systems which are perturbations of crystalline states remain tractable \textit{via} the aforementioned methods.

In this study, we present exact analytical results for the displacements of particles and the resulting force fields upon introduction of a small disorder into three dimensional crystalline states. We also describe the coarse-grained displacement correlation functions averaged over the quenched disorder. Most significantly, we show that the distribution of interparticle forces is a generalized Irwin-Hall distribution which captures crucial deviations from Gaussian behaviour at the tails of the distribution. We employ the analytical framework developed in references~\cite{acharya2020athermal,das2021long,acharya2021disorder}, by introducing disorder to an FCC crystal composed of soft, repulsive harmonic spheres, and making theoretical predictions of the microscopic structure of the system. We further validate our results through numerical simulations. We also outline the generalisation of these methods to higher dimensions.

\section{Model}

\begin{figure}[t]
\centering
\includegraphics[scale=0.6]{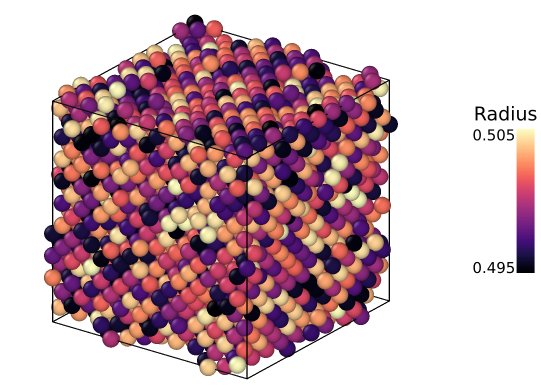}
\caption{An energy minimized configuration of $4000$ particles from an initial Face Centred Cubic (FCC) lattice. The gradient in color corresponds to the radii of the particles. As a result of the polydispersity in particle sizes, the system settles into a force balanced configuration deviating from the FCC arrangement. We treat the dispersion in sizes as a perturbation and study the corresponding changes in the interparticle forces and positions as a response.}
\end{figure}

We consider a system of particles arranged in an $L\times L \times L$ lattice, with periodic boundary conditions. For the FCC arrangement only the alternating lattice points are occupied by particles.  A neighbouring pair of particles $i$ and $j$ at positions $r_{i}$ and $r_{j}$ interact with a potential of the form
\begin{equation}
\begin{aligned}
V_{\{\sigma_{i j},K_{ij}\}}\left(\vec{r}_{i j}\right) &=K_{ij}\left(1-\frac{\left|\vec{r}_{i j}\right|}{\sigma_{i j}}\right)^{\alpha} & \text { for } \left|\vec{r}_{i j}\right|<\sigma_{i j} \\
&=0 & \text { for } \left|\vec{r}_{i j}\right| \geq \sigma_{i j},
\end{aligned}
\label{potential}
\end{equation}
where  $\vec{r}_{i j}=\vec{r}_{j}-\vec{r}_{i}$ represents the distance between the pair. The radii of the $i^{th}$ and $j^{th}$ particle are denoted as $\sigma_{i}$ and $\sigma_{j}$ respectively and the interaction cut-off distance between them, $\sigma_{i j}=\left(\sigma_{i}+\sigma_{j}\right)$. Here $K_{ij}$ denotes the stiffness of interaction between particles. 
We restrict our study to systems described by pairwise additive interactions. Moreover, since we only consider athermal states, the Hamiltonian does not contain terms corresponding to the kinetic energy and may be expressed as
\begin{equation}
H=\sum_{\langle ij\rangle} V\left(\vec{r}_{i j};\alpha,\sigma_{i j},K_{ij}\right).
\label{Hamiltonian}
\end{equation}

Such a Hamiltonian and the interaction potential may be used to model a wide variety of systems. The behaviour of paradigmatic soft sphere systems~\cite{das2021long,tong2015crystals,acharya2020athermal,acharya2021disorder} modelling frictionless materials such as foams and colloids may be recovered by using $\alpha \in \{3/2, 2, 5/2\}$ and $K_{ij} \equiv K$. Crystalline networks with random bond stiffness ~\cite{schirmacher1998harmonic,domb1959vibration,taraskin2001origin,gratale2013phonons} may also be modelled using a distribution of stiffness $P\left(K_{ij}\right)$, while retaining constant particle sizes, $\sigma_{i} \equiv \sigma$.

The initial arrangement consists of particles with identical radii $\sigma_{i} \equiv \sigma_{0}$ positioned on a crystalline lattice with the location of each particle denoted by $\{\vec{r}_i^{(0)}\} \equiv \{x_i^{(0)},y_i^{(0)},z_i^{(0)}\}$. The lattice constant of the system may be expressed as $R_{0} = 2 \sigma_{0}(1-\varepsilon)$, where $\varepsilon$ quantifies the overcompression or the overlap between two neighbouring particles in the initial crystalline state given by $2 \sigma_0\varepsilon=2\sigma_0\left(1-(\phi_{c} / \phi)^{1/d}\right)$. Here, $d$ is the spatial dimension, $\phi$ is the packing fraction of the system and $\phi_{c}$ is the packing fraction of the close packed crystal with no overlaps between the particles ($\varepsilon=0$). In the case of a three dimensional system, $\phi_{c} = \pi \sqrt{2} /6$. Numerical simulations are performed using Harmonic interparticle interactions $(\alpha = 2)$ with identical spring constants $(K_{ij} \equiv 1/2)$. The interparticle forces in the system are determined from  the potential given in equation~\eqref{potential} as
\begin{equation}
    \vec{f}_{i j}=\frac{\alpha K}{\sigma_{i j}}\left(1-\frac{\left|r_{i j}\right|}{\sigma_{i j}}\right)^{\alpha-1} \hat{r}_{i j},
\end{equation}
where the three components of the forces can be expressed as
\begin{equation}
f_{i j}^{\mu}=\frac{\alpha K}{\sigma_{i j}}\left(1-\frac{\left|\vec{r}_{i j}\right|}{\sigma_{i j}}\right)^{\alpha-1} \frac{r^{\mu}_{i j}}{\left|\vec{r}_{i j}\right|}.
\label{force_componenet}
\end{equation}

Here,  $r^{\mu}_{i}$ represents the three spatial components of position vector, $\{ x_i,y_i, z_i\}$, and $f^{\mu}_{ij}$ represents the three force components $\{ f^x_{ij},f^y_{ij}, f^z_{ij}\}$. We note that  equations for the force components are non-linear in $r^{\mu}_{i j}$ and $\sigma_{ij}$. For a crystalline ordered state of an over-compressed monodisperse system, we can rewrite these force components as

\begin{equation}
    f_{i j}^{\mu(0)}=\frac{\alpha K}{2\sigma_0}\left(1-\frac{R_0}{2\sigma_0}\right)^{\alpha-1}\hat{r}_{ij}^{(0)}.
\end{equation}

\begin{figure}[t]
\centering
\includegraphics[scale=0.43]{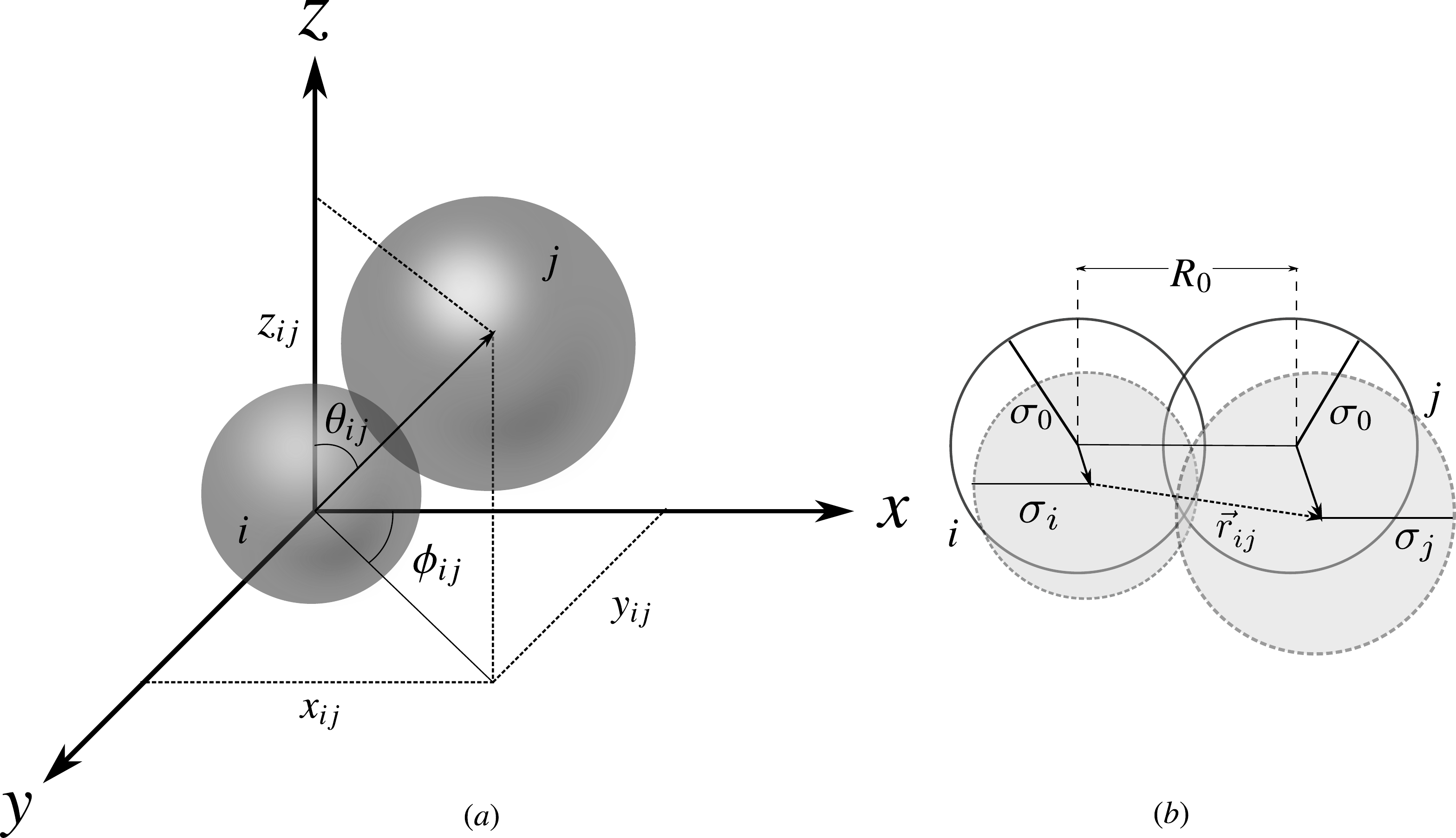}
\caption{Schematic representation of two neighbouring particles in an FCC lattice. In figure~(a), we represent all the Cartesian components of the relative distance vector $\vec{r}_{ij}$ between the $i^{th}$ and $j^{th}$ particle. We have also shown both the polar $(\theta_{ij})$ and azimuth $(\phi_{ij})$ angles made by this vector with the coordinate axes.  In figure (b), we represent both the particle sizes as well as their positions before and after the introduction of particle size disorder. }
\label{schematic spherical cordinates}
\end{figure}

In the above equation $R_0$ is the initial lattice spacing whereas  $\sigma_{i}^{(0)}=\sigma_0$.  This relative distance vector between two neighbouring particles in the initial crystal is represented as $\vec{r}_{ij}^{(0)}=R_0 \hat{r}_{ij}^{(0)}$, where $\hat{r}_{ij}^{(0)}$ are the unit vectors along the basis direction of the underlying lattice. These unit vectors can be written in spherical coordinates as (see ~\fref{schematic spherical cordinates})
\begin{equation}
    \hat{r}_{ij}^{(0)}= \cos(\phi_{ij}^{(0)})\sin(\theta_{ij}^{(0)}) \hat{x}+\sin(\phi_{ij}^{(0)})\sin(\theta_{ij}^{(0)}) \hat{y} + \cos(\theta_{ij}^{(0)})\hat{z}.
    \label{unit vector}
\end{equation}

\section{Linearized perturbation expansion of the displacement and force fields}

Following the framework developed in references~\cite{acharya2020athermal,das2021long,acharya2021disorder} for two dimensional systems, we study the displacements of particles from their crystalline lattice positions, incurred upon introducing a disorder in the radii ($\sigma_{i}$) of individual particles. We also derive the effects on the interparticle force fields in the material. The disorder in particle sizes is introduced \textit{via} a polydispersity which represents a perturbation to the initial sizes
\begin{equation}
    \sigma_{i}=\sigma_{0}+\delta \sigma_{i}=\sigma_{0}+\sigma_{0}\eta \zeta_{i}.
\end{equation}

Here, $\eta$ is a parameter that controls the strength of disorder and $\zeta_i$ is a random number sampled from a  uniform distribution spanning $[-1/2, 1/2]$. We therefore have
\begin{equation}
    \left\langle \delta \sigma_{i} \delta \sigma_{j} \right\rangle= \frac{\eta^2}{48} \delta_{ij}.
    \label{sigma_sigma_avg}
\end{equation}

As a response to this disorder, in order to maintain force balance, the particles move from their crystalline positions
\begin{equation}
  r^{\mu}_{i}=r^{\mu(0)}_{i}+\delta r^{\mu}_{i}.
\end{equation}
It is also convenient to define relative displacements between two neighboring particles 
\begin{equation}
\delta r^{\mu}_{ij}= \delta r^{\mu}_{j}-\delta r^{\mu}_{i}.
\end{equation}
Crucially, the perturbation in the radii also changes the sum of the radii $\sigma_{ij} = \sigma_{ij}^{(0)} +\delta \sigma_{ij}$, which represents the interaction cut-off between neighbouring particles, with
\begin{equation}
\delta \sigma_{i j}= \delta \sigma_{i}+\delta \sigma_{j}.
\end{equation}

As a response to the change in the positions of the particles, there is a corresponding change in the forces $\vec{f}_{ij}$ between any two neighbouring particles $i$ and $j$. This can be written as a perturbation about the initial force between the particles as, 
\begin{equation}
    f_{i j}^{\mu}= f_{i j}^{\mu(0)}+\delta f_{i j}^{\mu}.
\end{equation}

Now each component of the excess force $\delta f_{ij}^{\mu}$ between particles $i$ and $j$ can be Taylor expanded about its value in the crystalline ground state, in terms of $\delta r^{\mu}_{ij}$ and $\delta \sigma_{ij}$ as
\begin{equation}
\delta f_{i j}^{\mu}=\sum_{\nu}C_{i j}^{\mu \nu} \delta r^{\nu}_{i j}+C_{i j}^{\mu \sigma} \delta \sigma_{i j} +\sum_{\nu}\sum_{\gamma}C_{i j}^{\mu \nu \gamma} \delta r^{\nu}_{i j}\delta r^{\gamma}_{i j}
+\sum_{\nu}C_{i j}^{\mu \nu \sigma} \delta r^{\nu}_{i j}\delta \sigma_{i j}
+C_{i j}^{\mu \sigma \sigma} \delta \sigma_{i j} \delta \sigma_{i j}+\mathcal{O}(\delta^3).
\label{net force}
\end{equation}
where,
\begin{equation}
   \left. C_{ij}^{\mu \nu}= \frac{\partial f_{ij}^{\mu}}{\partial r_{ij}^{ \nu}}\right|_{\vec{r}_{ij},\sigma_{ij}=\vec{r}_{ij}^{(0)},\sigma_{ij}^{(0)}}, \hspace{0.5cm}\left. C_{ij}^{\mu \nu \gamma}= \frac{\partial^2 f_{ij}^{\mu}}{\partial r_{ij}^{\nu} \partial r_{ij}^{\gamma}}\right|_{\vec{r}_{ij},\sigma_{ij}=\vec{r}_{ij}^{(0)},\sigma_{ij}^{(0)}},~\ldots
\end{equation}

Since the coefficients $C_{ij}^{\mu \nu}, C_{ij}^{\mu \nu \gamma},~\ldots$ only depend on the initial crystalline state that is translationally invariant, the coefficients themselves do not depend on the particle indices $i$ and $j$, but only on the orientation of the contact between them, represented by the polar angles ($\theta_{ij}, \phi_{ij}$). For example in an FCC lattice there are twelve such coefficients corresponding to the twelve different bonds for each particle. For a system where particles interact via potentials given in equation~\eqref{potential}, with $K_{ij}=K$, the coefficients at first order corresponding to a perturbation in the radii can be written as
\begin{equation}
\begin{aligned}
    & C^{\mu \nu}_{ij} =\frac{-\alpha K (1 - \tilde{R}_0)^{-2 + \alpha}}{4\tilde{R}_0\sigma_0^2}\left[  \left( 1 + \tilde{R}_0 (-2 + \alpha) \right) r_{ij}^{\mu(0)}. r_{ij}^{\nu(0)} - ( 1-\tilde{R}_0 )\delta_{\mu , \nu}\right],\\
    & C^{\mu \sigma}_{ij} =\frac{\alpha K (1 - \tilde{R}_0)^{-2 + \alpha}}{4\sigma_0^2} (1 - \alpha \tilde{R}_0)r_{ij}^{\mu(0)},
\end{aligned}
\label{coefficients}
\end{equation}

where $\tilde{R}_0 = R_0/2\sigma_0$. Next, given a perturbation in particle sizes of the form $\sigma_i=\sigma_0 +\lambda \delta \sigma_i$, we may expand the resultant positions order-by-order in terms of the tuning parameter $\lambda$ which will be set to 1 at the end of the calculation   

\begin{equation}
\delta r_{i}^{\mu} = \lambda \delta r^{\mu(1)}_{i}+\lambda^2 \delta r^{\mu(2)}_{i}+\lambda^3 \delta r^{\mu(3)}_{i}+.... 
\label{ordered disp}
\end{equation}

Here $ \delta r^{\mu(n)}_{i} \equiv \{\delta x_i^{(n)},\delta y_i^{(n)},\delta z_i^{(n)}\}$ represent the $n^{th}$ order displacement fields. We also define the relative displacements at every order as, $\delta r^{\mu(n)}_{ij} = \delta r^{\mu(n)}_{j}-\ \delta r^{\mu(n)}_{i}$.
Similarly, the change in forces between two neighbouring particles given in equation~\eqref{net force} can also be expressed as an expansion in the parameter $\lambda$, given by
\begin{equation}
    \delta f_{i j}^{\mu}=\lambda \delta f_{i j}^{\mu(1)}+\lambda^{2} \delta f_{i j}^{\mu(2)}+\lambda^{3} \delta f_{i j}^{\mu(3)}+\mathcal{O}\left(\lambda^{4}\right).
    \label{ordered force}
\end{equation}

Inserting the expressions for the excess inter-particle forces $\delta f_{ij}$ given by equation~\eqref{ordered force} and relative displacements given by  equation~\eqref{ordered disp} into the Taylor expansion of the excess forces, $\delta f_{ij}$ in  equation~\eqref{net force},
and matching terms at first order in $\lambda$, we obtain the change in interparticle forces to first order  as,
\begin{equation}
    \delta f_{i j}^{\mu(1)}=\sum_{\nu} C_{i j}^{\mu \nu} \delta r^{\nu(1)}_{i j}+C_{i j}^{\mu \sigma} \delta \sigma_{i j}.
    \label{df linear}
\end{equation}

Similarly, matching terms at second and higher order of $\lambda$ in equation~\eqref{net force}, we arrive at the higher order correction to the changes in the inter-particle forces. For example the second order correction is given by,
\begin{equation}
\delta f_{i j}^{\mu(2)}= \sum_{\nu} C_{i j}^{\mu \nu} \delta r^{\nu(2)}_{i j} +\sum_{\nu}\sum_{\gamma}C_{i j}^{\mu \nu \gamma} \delta r^{\nu(1)}_{i j}\delta r^{\gamma(1)}_{i j}+\sum_{\nu}C_{i j}^{\mu \nu \sigma} \delta r^{\nu(1)}_{i j}\delta \sigma_{i j} +C_{i j}^{\mu \sigma \sigma} \delta \sigma_{i j} \delta \sigma_{i j}.
\end{equation}

\subsection{Green's function solutions to the displacement and force fields}
 
Since we are considering systems where frictionless particles interact only \textit{via} pairwise central potentials, there are no microscopic torques in the system. In an energy minimized configuration the net force on each particle is zero as the system is in a stable equilibrium. Such a force balance on every grain $i$ can be expressed as
\begin{equation}
  \sum_{j=0}^{z}\vec{f}_{i j}=0.
\end{equation}

Here $z$ is the coordination number of the underlying crystalline lattice, which remains unchanged in the regimes of small disorder in particle sizes (pre contact breaking state)~\cite{maharana2022first}.  We begin by imposing the force-balance constraints to the first
order terms in our expansion of the inter-particle forces, 
\begin{equation}
\begin{aligned}
&\sum_{j=0}^{z} f^{\mu}_{ij}= \sum_{j=0}^{z} \left(f^{\mu(0)}_{ij} +\delta f_{i j}^{\mu(1)}\right) =0,\\
&\sum_{j=0}^{z} \left(\sum_{\nu}C_{i j}^{\mu \nu}\delta r^{\nu(1)}_{ij} +C_{i j}^{\mu \sigma} \delta \sigma_{i j} \right)=0 ,
\end{aligned}   
\label{force balance linear}
\end{equation}

where the coefficients $C_{ij}^{\mu \nu}$ depend only on the lattice constant $R_0$ and the orientation of the neighbour $j$ with respect to  particle $i$, but are independent of the particle index $i$. Since every particle in the system has $3$ degrees of freedom in three dimensions, there are $3N$ force-balance constraints at every order for an $N$-particle system. Given such a set of coupled linear equations, we need to invert a $3N \times 3N$ matrix to obtain the displacement field at any order, for a given realisation of the quenched disorder. Since it is computationally expensive to invert large matrices corresponding to large system sizes, we need to find an efficient transformation that will diagonalize the matrix. Here the coefficients $C_{ij}^{\alpha \beta}$ in the linear expansion are translationally invariant, and therefore a discrete Fourier transform of the above equations will block-diagonalize the matrix into blocks of size $3 \times 3$ corresponding to each reciprocal lattice point $\vec{k}$. The Fourier transforms of the
changes in positions and radii can be defined as
\begin{equation}
    \begin{aligned}
    &\delta \tilde{r}^{\mu(1)}(\vec{k})=\sum_{\vec{r}} \exp (i \vec{k} \cdot \vec{r}) \delta r^{\mu(1)}(\vec{r}) \\
    &\delta \tilde{\sigma}(\vec{k})=\sum_{\vec{r}} \exp (i \vec{k}. \vec{r}) \delta \sigma(\vec{r}).
    \end{aligned}
    \label{disp1st}
\end{equation}
Here $\vec{r}$ labels the sites of the lattice and $\vec{k}$ represents the reciprocal lattice vectors defined as $k_\mu=2\pi/L_\mu$, where $L_\mu$ is the number of grid points along the $\mu$-axis ($\mu =x,y,z$). We note that the discrete Fourier transform defined above involves the original crystalline positions $\vec{r} \equiv \vec{r}^{(0)}$, which we use interchangeably with the site index, $\vec{r} \equiv i$. It is also convenient to define the basic translation coefficients in Fourier space as 
\begin{equation}
    \mathcal{F}_{j}(\vec{k}) = \exp (-i \vec{k}. \vec{r}) .
\end{equation}

Multiplying equation~\eqref{force balance linear} by $\exp (i \vec{k}. \vec{r})$ and summing over all the lattice sites $\vec{r} \equiv i$ leads to the following matrix equation at every $\vec{k}$,
\begin{equation}
\underbrace{\left(\begin{array}{ccc}
A^{x x}(\vec{k}) & A^{x y}(\vec{k}) & A^{x z}(\vec{k}) \\
A^{y x}(\vec{k}) & A^{y y}(\vec{k}) & A^{y z}(\vec{k}) \\
A^{z x}(\vec{k}) & A^{z y}(\vec{k}) & A^{z z}(\vec{k})
\end{array}\right)}_{\textbf{A}}\left(\begin{array}{c}
\delta \tilde{x}(\vec{k}) \\
\delta \tilde{y}(\vec{k}) \\
\delta \tilde{z}(\vec{k})
\end{array}\right)=\delta \sigma(\vec{k})\left(\begin{array}{c}
D^{x}(\vec{k}) \\
D^{y}(\vec{k}) \\
D^{z}(\vec{k}) 
\end{array}\right),
\label{mateqF}
\end{equation}

where,
\begin{equation}
\begin{aligned}
    & A^{\alpha \beta}(\vec{k})=\sum_{j=0}^{z} \left[1-\mathcal{F}_{j}(\vec{k})\right] C_{i j}^{\alpha \beta}(\theta_j,\phi_j),\\
    & D^{\alpha }(\vec{k})=-\sum_{j=0}^{z} \left[1+\mathcal{F}_{j}(\vec{k})\right] C_{i j}^{\alpha \sigma}(\theta_j,\phi_j).
\end{aligned}
\label{Amat}
\end{equation}

All the matrix elements in the above equation can be expressed using the coefficients $C_{ij}^{\alpha \beta}$, which depend only on the underlying lattice.
We may interpret the matrix $\bold{A}$ as being an inverse Green’s function of the system in Fourier space. Therefore, the Green's function in Fourier space may in-turn be interpreted as the inverse of matrix $\bold{A}$ and can be expressed as:
\begin{equation}
\bold{\tilde{\mathcal{G}}}=\bold{A}^{-1}=\left[\begin{array}{lll}
\tilde{\mathcal{G}}^{x x}(\vec{k}) & \tilde{\mathcal{G}}^{x y}(\vec{k}) & \tilde{\mathcal{G}}^{x z}(\vec{k})\\
\tilde{\mathcal{G}}^{y x}(\vec{k}) & \tilde{\mathcal{G}}^{y y}(\vec{k}) & \tilde{\mathcal{G}}^{y z}(\vec{k}) \\
\tilde{\mathcal{G}}^{z x}(\vec{k}) & \tilde{\mathcal{G}}^{z y}(\vec{k}) & \tilde{\mathcal{G}}^{z z}(\vec{k}) 
\end{array}\right].
\label{Greenk matrix}
\end{equation}

The elements $\tilde{\mathcal{G}}^{\mu v}$ can be expressed in terms of the elements, $A^{\mu v}$ of the matrix $\bold{A}$, which depend on the underlying potential between the particles through the coefficients $C_{i j}^{\alpha \beta}$. 
Now inverting the matrix $\bold{A}$ in equation~\eqref{mateqF} leads to an expression for the displacement fields in Fourier space to first order, in terms of the Fourier transform of the quenched disorder $\{\delta \tilde{\sigma}(\vec{k})\}$. We have
\begin{equation}
    \delta \tilde{r}^{\mu}(\vec{k})=\delta\tilde{\sigma}(\vec{k})\tilde{G}^{\mu}(\vec{k}),
    \label{dispk}
\end{equation}
where $\tilde{G}^{\mu}(\vec{k})$ represents the Fourier transform of the Green's function for the displacement fields in real space which can be written as
\begin{equation}
    \tilde{G}^{\mu}(\vec{k})=\sum_{\nu}\tilde{\mathcal{G}}^{\mu \nu}(\vec{k}) D^{\nu}(\vec{k}).
\label{greenk}
\end{equation}

Finally the inverse Fourier transform of equation~\eqref{dispk}, which is the convolution of the inverse Fourier transform of $\tilde{G}^{\mu}(\vec{k})$ and $\tilde{\delta\sigma}(\vec{k})$, yields the displacement fields in real space up to the first order in the perturbation expansion, which is represented as,
\begin{equation}
\delta r^{\mu(1)}(\vec{r})=\sum_{r^{\prime}} G^{\mu}\left(\vec{r}-\vec{r}^{\prime}\right) \delta \sigma\left(\vec{r}^{\prime}\right).
\label{dr linear}
\end{equation}
Above the Green's function for the displacement fields can be written as,
\begin{equation}
G^{\mu }(\vec{r})=\frac{1}{V} \sum_{\vec{k}} e^{-i \vec{k}. \vec{r}} \tilde{G}^{\mu }(\vec{k}).
\end{equation}

This Green's function can also be interpreted  as the linear order displacement field at lattice site $\vec{r}$ due to a unit particle size disorder at the origin. The above formalism that yields the displacement fields can be generalized to $d$-dimensions given the Green's functions for a $d$-dimensional solid. In the thermodynamic limit $(L  \to \infty)$, the volume occupied by a single mode in $k$-space i.e. $(2\pi/L)^d $ tends to zero, and we can transform the summation in the above equation into an integral as
\begin{equation}
    G^{\mu }(\vec{r})=\frac{1}{(2\pi)^d}\int_{-\pi}^{\pi}d^dk  e^{-i \vec{k}.\vec{ r}} \tilde{G}^{\mu }(\vec{k}).
\end{equation}

Now we can use the linear displacement fields derived above in equation~\eqref{df linear} to obtain the excess linear force between two neighbouring particles as a linear function of the particle size disorders,
\begin{equation}
\begin{aligned}
    \delta f_{i j}&^{\mu(1)}= \sum_{r^{\prime}}\left[\sum_{\nu} C_{i j}^{\mu \nu} \left(G^{\nu}\left(\vec{r}_j-\vec{r}^{\prime}\right)-G^{\nu}\left(\vec{r}_i-\vec{r}^{\prime}\right)\right)\right]  \delta \sigma\left(\vec{r}^{\prime}\right)+C_{i j}^{\mu \sigma}\delta \sigma_{i j} \\
    &=\sum_{r^{\prime}}\left[\sum_{\nu} C_{i j}^{\mu \nu} \left(G^{\nu}\left(\vec{r}_j-\vec{r}^{\prime}\right)-G^{\nu}\left(\vec{r}_i-\vec{r}^{\prime}\right)\right)+ C_{i j}^{\mu \sigma}\left(\delta(\vec{r}_i-\vec{r}^{\prime})+\delta(\vec{r}_j-\vec{r}^{\prime}) \right)\right]  \delta \sigma\left(\vec{r}^{\prime}\right).
\end{aligned}
\label{eq_excess_force_components}
\end{equation}

\subsection{Distributions of displacement and force fields}

Next, we derive the distributions of several fluctuating quantities in the system due to the introduction of disorder. For a small amount of disorder, since the fluctuations are well approximated by a first order perturbation expansion, they are linearly proportional to the particle size disorder at every site as may be seen in equation~\eqref{eq_excess_force_components}. Since the disorder parameter, $\zeta$ is a uniform random number between $[-1/2,+1/2]$, the fluctuating quantities such as the contact forces and the displacements at every lattice site governed by a generalized Irwin-Hall distribution~\cite{marengo2017geometric, batsyn2013analytical} derived in~\ref{appendix:B}. This distribution also appears in the computation of the first contact breaking statistics in this model~\cite{maharana2022first}. These distributions are approximately Gaussian as has been observed in previous studies~\cite{acharya2020athermal}. However they deviate at the tails (see ~\fref{fig_fparallel}~(b)). The extremes of these distributions have been shown to determine the distributions of contact breaking events in the system~\cite{maharana2022first}, and so utilising the exact Irwin-Hall distributions is necessary in the configurations close to each such events.

In order to illustrate the effects of disorder, we focus on the distributions of forces between the particles. As we show below, that the forces in a crystal are primarily confined along the lattice directions. Therefore a convenient decomposition of the resultant forces is along and perpendicular to lattice directions i.e. $\vec{f}_{ij}=\vec{f}_{ij}^{\parallel}+\vec{f}_{ij}^{\perp}$ where,

\begin{equation}
\begin{aligned}
    &\vec{f}_{ij}^{\parallel}=\left(f_0 +\delta f_{ij}^{||}\right) \hat{r}_{ij}^{(0)}, \\
    & \vec{f}_{ij}^{\perp}= \vec{\delta f}_{ij}^{\perp}.
\end{aligned}
\end{equation} 

The component of the excess force parallel to the initial bond direction is the scalar product of the excess force between the neighbouring particles and the unit vector along the bond in the initial crystal structure,
\begin{equation}
\begin{aligned}
    \delta f_{i j}^{||}= \sum_{\mu}  \delta f_{i j}^{\mu(1)} r_{ij}^{\mu(0)}=\sum_{\vec{r}^{\prime}} C_{ij}^{||}(\vec{r}_i-\vec{r}^{\prime})\delta \sigma\left(\vec{r}^{\prime}\right)=\eta\sigma_0\sum_{\vec{r}^{\prime}} C_{ij}^{||}(\vec{r}_i-\vec{r}^{\prime})\zeta\left(\vec{r}^{\prime}\right),
\end{aligned}
\label{eq_fparallel}
\end{equation}
where,
\begin{equation}
\begin{aligned}
    C_{ij}^{||}(\vec{r}_i-\vec{r}^{\prime})=\sum_{\mu}\sum_{\nu}  &r_{ij}^{\mu(0)}\left[C_{i j}^{\mu \nu} \left(G^{\nu}\left(\vec{r}_j-\vec{r}^{\prime}\right)-G^{\nu}\left(\vec{r}_i-\vec{r}^{\prime}\right)\right)\right.\\&
    \left. + C_{i j}^{\mu \sigma}\left(\delta(\vec{r}_i-\vec{r}^{\prime})+\delta(\vec{r}_j-\vec{r}^{\prime}) \right)\right].
\end{aligned}
\end{equation}

The perpendicular component of the excess force lies on a plane perpendicular to the original crystalline directions and may be easily expressed as, $\vec{\delta f}_{ij}^{\perp}= \vec{f}_{ij}-\vec{\delta f}_{ij}^{\parallel}$. As  $ \delta f_{i j}^{\parallel}$ is defined as a linear summation of uniform random variables ($\zeta_i$) in equation~\eqref{eq_fparallel} with multiplicative coefficients, the distribution of $ \delta f_{i j}^{\parallel}$ is a generalized form of the Irwin-Hall distribution, as derived in~\ref{appendix:B}.  The cumulative distribution function (CDF) of $ \delta f_{i j}^{||}$ for a system of $N$ particles is given by
\begin{equation}
\begin{aligned}
     &F\left(\delta f_{i j}^{||}\right)=\frac{1}{N!\prod_{m=1}^{N}C_{ij}^{||}(\vec{r}_i-\vec{r}_m) }\left[\left(\frac{\delta f_{i j}^{||}}{\eta \sigma_0}+\frac{1}{2}\sum_{m=1}^{N}C_{ij}^{||}(\vec{r}_i-\vec{r}_m)\right)^{N}-\right.\\
     &\left.\sum_{k=1}^{N}(-1)^{k-1}\sum_{1 \leq j_{1}<j_{2}<\cdots<j_{k} \leq N} 
f\left(\frac{\delta f_{i j}^{||}}{\eta \sigma_0}+\frac{1}{2}\sum_{m=1}^{N}C_{ij}^{||}(\vec{r}_i-\vec{r}_m)-\sum_{n=1}^{k}C_{ij}^{||}(\vec{r}_i-\vec{r}_{j_n})\right)^N\right],
\label{eq_irwin1}
\end{aligned}
\end{equation}
where $f(x)=x\Theta(x)$ and $\Theta$ is the Heaviside theta function. The PDF may then be expressed as
\begin{equation}
    P(\delta f_{i j}^{||})=\left|\frac{d F(\delta f_{i j}^{||})}{d \delta f_{i j}^{||}}\right|.
    \label{eq_irwin2}
\end{equation}

Since $f_{ij}^{\parallel}=f_0+\delta f_{ij}^{\parallel}$ , the parallel component of the force ($f_{ij}^{\parallel}$) will possess the same distribution as $\delta f_{ij}^{\parallel}$ shifted by an amount $f_0$. Similarly we can derive expressions for the distribution of displacement fields or different components of the force fields as they can also be written as linear summations of uniform random variables with multiplicative coefficients.

\subsection{Displacement correlations in Real space}

Next, we derive the correlations between the  displacements of particles at various lattice sites  induced by the disorders in the particle sizes.
The correlations in the displacement fields in real space between the particles at two lattice points $\vec{r}$ and $\vec{r}'$ can be represented as 

\begin{equation}
  \mathcal{C}_{\mu \nu}\left(\vec{r}-\vec{r}^{\prime}\right)=\left\langle\delta r_{\mu}(\vec{r}) \delta r_{\nu}\left(\vec{r}^{\prime}\right)\right\rangle.
\end{equation}

Here the angular brackets $\left\langle\ldots\right\rangle$ denote an average over the quenched disorder.  Now we can replace the  displacements in real space by their Fourier transforms in the above equation as, 
\begin{equation}
\begin{aligned}
    \left\langle\delta r_{\mu}(\vec{r}) \delta r_{\nu}(\vec{r}^{\prime})\right\rangle=&\left\langle\frac{1}{V^{2}} \sum_{\vec{k}, \vec{k}^{\prime}} \delta \tilde{r}_{\mu}(\vec{k}) \delta \tilde{r}_{\nu}(\vec{k}^{\prime}) \exp (-i \vec{k} \cdot \vec{r}) \exp \left(-i \vec{k}^{\prime} \cdot \vec{r}^{\prime}\right)\right\rangle.\\
    =&\frac{1}{V^{2}} \sum_{\vec{k}, \vec{k}^{\prime}}\left\langle \delta \tilde{r}_{\mu}(\vec{k}) \delta \tilde{r}_{\nu}(\vec{k}^{\prime})\right\rangle \exp (-i \left(\vec{k} \cdot \vec{r}+\vec{k}^{\prime} \cdot \vec{r}^{\prime}\right)) .
\end{aligned}
    \label{disp_cor}
\end{equation}

Upon inserting the expression for $\delta r_{\mu}(\vec{k})$ provided in equation~\eqref{dispk} in the above expression, we obtain the linear order displacement correlations in real space, in terms of the Fourier transform of the particle size disorder corresponding to different wave vectors, $\vec{k}$. i.e., the term inside the angular brackets in equation~\eqref{disp_cor} resolves to
\begin{equation}
 \left\langle\delta r_{\mu}(\vec{k}) \delta r_{\nu}(\vec{k}^{\prime})\right\rangle= \tilde{G}^{\mu }(\vec{k}) \tilde{G}^{\nu }(\vec{k}^{\prime})\left\langle\delta \tilde{\sigma}(\vec{k}) \delta \tilde{\sigma}(\vec{k}^{\prime})\right\rangle. 
 \label{disp_cork}
\end{equation}

Since the system is translationally invariant, the correlation function does not depend on the actual positions of the particles, but only on the separation between them. This implies that a translation of the system by any distance, say $\Delta \vec{r}$, should leave the correlation function in the equation~\eqref{disp_cor}, invariant. However, performing such a translation yields an extra term $\exp\left(-i (\vec{k}+\vec{k}')\cdot\Delta \vec{r}\right)$. Therefore, in order to satisfy translational invariance, the expressions in equation~\eqref{disp_cork} is non-zero only when $\vec{k}=-\vec{k}^{\prime}.$ Thus, we have 
\begin{equation}
\begin{aligned}
    \left\langle\delta \tilde{\sigma}(\vec{k}) \delta \tilde{\sigma}(-\vec{k})\right\rangle &=\sum_{\vec{r},\vec{r}^{\prime}}  \left\langle\delta \sigma(\vec{r}) \delta \sigma\left(\vec{r}^{\prime}\right)\right\rangle \exp (i \vec{k} \cdot (\vec{r}-\vec{r}^{\prime})) \\
    &=\frac{\eta^{2}}{48} \sum_{\vec{r},\vec{r}^{\prime}}  \delta\left(\vec{r}-\vec{r}^{\prime}\right)\exp (i \vec{k} \cdot (\vec{r}-\vec{r}^{\prime}))=\frac{\eta^{2}}{48} V.
\end{aligned}
\end{equation}
Above we have used $\left\langle \delta \sigma_{i} \delta \sigma_{j} \right\rangle= \frac{\eta^2}{48} \delta_{ij}$ as given in equation~\eqref{sigma_sigma_avg}. Now the real space displacement correlation takes the form,
\begin{equation}
  \mathcal{C}_{\mu \nu}\left(\vec{r}-\vec{r}^{\prime}\right)=\frac{\eta^{2}}{48 V} \sum_{\vec{k}}\tilde{G}^{\mu }(\vec{k}) \tilde{G}^{\nu }(-\vec{k})\exp (-i \vec{k} \cdot (\vec{r}-\vec{r}^{\prime})).
\end{equation}
In the $L  \to \infty$ limit, we can rewrite the sum in the above equation as an integral,
\begin{equation}
    \mathcal{C}_{\mu \nu}\left(\vec{r}-\vec{r}^{\prime}\right)=\frac{\eta^{2}}{48}\frac{1}{(2\pi)^d}\int d^dk \exp (-i \vec{k} \cdot (\vec{r}-\vec{r}^{\prime}))\tilde{G}^{\mu }(\vec{k}) \tilde{G}^{\nu }(-\vec{k}).
    \label{eq_disp_correlation}
\end{equation}

\section{Numerical simulations of an FCC arrangement of harmonic spheres}

To verify the predictions from our theory, we simulate an athermal, over-compressed FCC lattice  of soft friction-less particles into which we introduce disorder in particle sizes. Here we consider states with each particle in force balance, i.e. configurations at energy minima (there are no torques given that the particles are friction-less). We then minimize the energy of the system using the FIRE (Fast Inertial Relaxation Engine) algorithm ~\cite{bitzek2006structural}. In this study, we have chosen a packing fraction $\phi=0.80$ indicating the degree of compression when compared to the marginally jammed FCC packing fraction $\phi_{c} \approx 0.74$. The initial particle radius with no disorder is chosen to be $\sigma_0=0.5$, which, along with $\phi$ determines the interparticle distance in the reference crystal to be $R_0 \approx 0.9826116$. The value of bond stiffness is kept fixed at $K = 0.5$. The numerical results provided
in this paper are averaged over $1000$ realisations of disorder for a system size of $N = 4000$ for different strengths of particle size disorder ($\eta$).

\begin{figure}[t]
\centering
\includegraphics[scale=0.50]{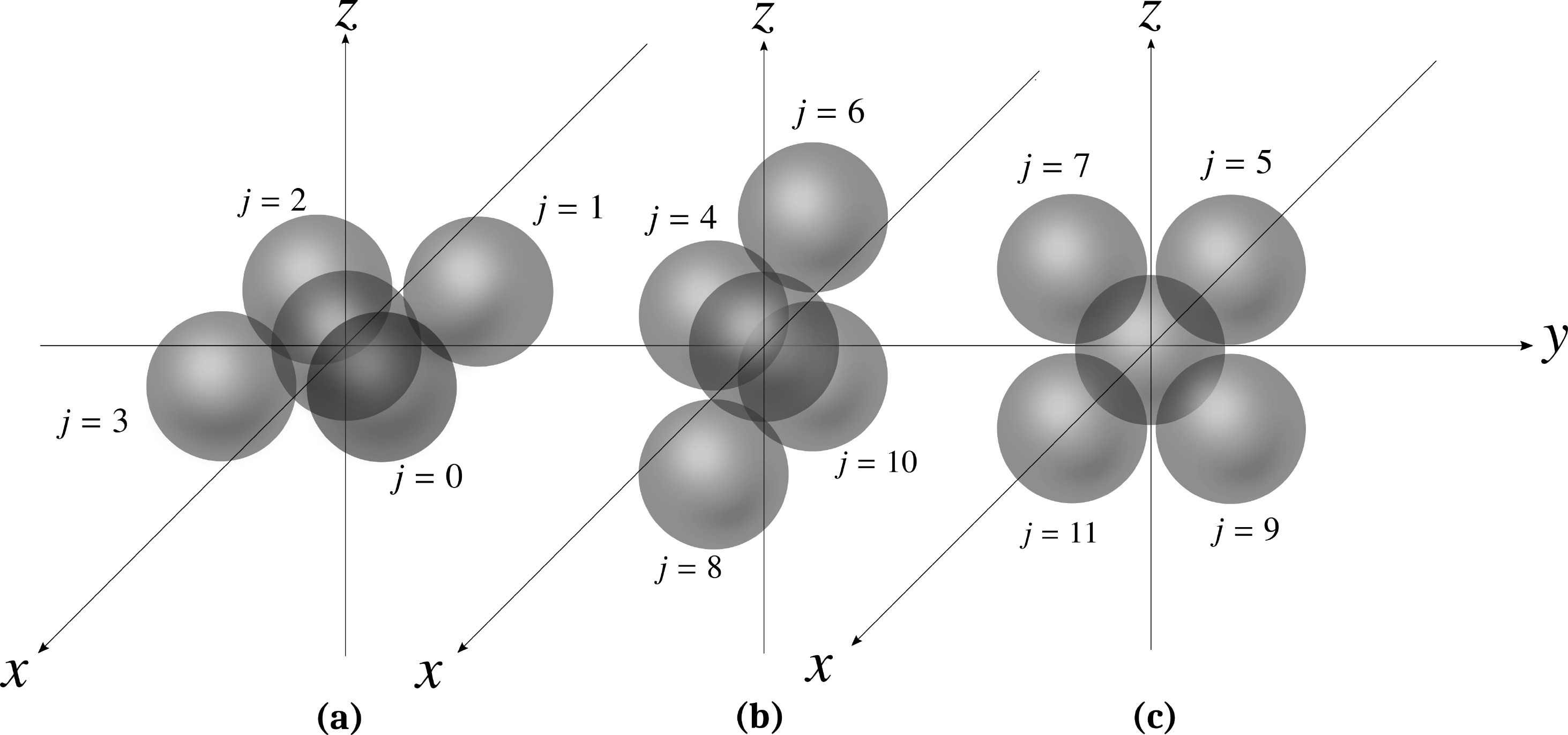}
\caption{Schematic representation of all the neighbouring particles (with neighbour index $j$) in an FCC lattice.  }
\label{fig_neighbours}
\end{figure}

In an FCC lattice, each particle has 12 neighbours all of which are at an equal distance $R_{0}$ in the initial crystalline state. We display a schematic diagram of all the neighbours in a Cartesian coordinate system in ~\fref{fig_neighbours}. To begin with, we consider a cubic lattice with a grid spacing of $R_{0}$. In order to generate an FCC arrangement, we place particles at alternating grid points. i.e., if the position of any particle in the lattice is represented as the number of grid points along each axis direction as $(n_x,n_y,n_z)$ then the condition for an FCC lattice is $n_x+n_y+n_z=2n$, where $n$ is an integer. The above method to generate an FCC lattice is an extension of the method used in reference~\cite{horiguchi1972lattice} for creating a triangular lattice. A cubic lattice contains a total of $8L^3$ grid points (since the number of grid points along every coordinate axes is $2L$), out of which only half of them are filled with particles i.e. $N=4L^3$. In a spherical coordinate system, each of these neighbours may be represented by a set of 2 angles $(\theta,\phi)$ with respect to the central particle. In an FCC lattice the azimuthal and polar angles of all the 12 neighbours with respect to the central particle are as follows

\begin{equation}
\begin{array}{l}
\Theta\left(\theta_{0}, \phi_{0}\right)=\left(\frac{\pi}{2}, \frac{\pi}{4}\right) ~~~~~~~~~~~~\Theta\left(\theta_{1}, \phi_{1}\right)=\left(\frac{\pi}{2}, \frac{3 \pi}{4}\right) \\
\Theta\left(\theta_{2}, \phi_{2}\right)=\left(\frac{\pi}{2}, \frac{5 \pi}{4}\right)~~~~~~~~~~~
\Theta\left(\theta_{3}, \phi_{3}\right)=\left(\frac{\pi}{2}, \frac{7 \pi}{4}\right) \\
\Theta\left(\theta_{4}, \phi_{4}\right)=\left(\frac{\pi}{4}, 0\right) ~~~~~~~~~~~~
\Theta\left(\theta_{5}, \phi_{5}\right)=\left(\frac{\pi}{4}, \frac{\pi}{2}\right) \\
\Theta\left(\theta_{6}, \phi_{6}\right)=\left(\frac{\pi}{4}, \pi\right) ~~~~~~~~~~~~
\Theta\left(\theta_{7}, \phi_{7}\right)=\left(\frac{\pi}{4}, \frac{3 \pi}{2}\right) \\
\Theta\left(\theta_{8}, \phi_{8}\right)=\left(\frac{3 \pi}{4}, 0\right) ~~~~~~~~~~~
\Theta\left(\theta_{9}, \phi_{9}\right)=\left(\frac{3 \pi}{4}, \frac{\pi}{2}\right) \\
\Theta\left(\theta_{10}, \phi_{10}\right)=\left(\frac{3 \pi}{4}, \pi\right)~~~~~~~~~
\Theta\left(\theta_{11}, \phi_{11}\right)=\left(\frac{3 \pi}{4}, \frac{3 \pi}{2}\right)
\end{array}
\label{angles}
\end{equation}

We may now use equations~\eqref{unit vector}, \eqref{coefficients} and \eqref{angles} to obtain the coefficients of the linear expansion of the forces between two neighbouring particles. This in turn can be used to derive the Green's function that relates the displacement field and the particle size disorder in Fourier space, as given in equation~\eqref{dispk}. Here, $\vec{r} \equiv (n_x,n_y,n_z)$ labels the lattice sites in the triangular lattice, and the grid points in reciprocal lattice are given as,
$$
\vec{k} \equiv\left(k_{x}, k_{y}, k_{z}\right) \equiv\left(\frac{2 \pi l}{2 L}, \frac{2 \pi m}{2 L},\frac{2 \pi p}{2 L}\right).
$$

\begin{figure}[t]
\centering
\includegraphics[scale=0.62]{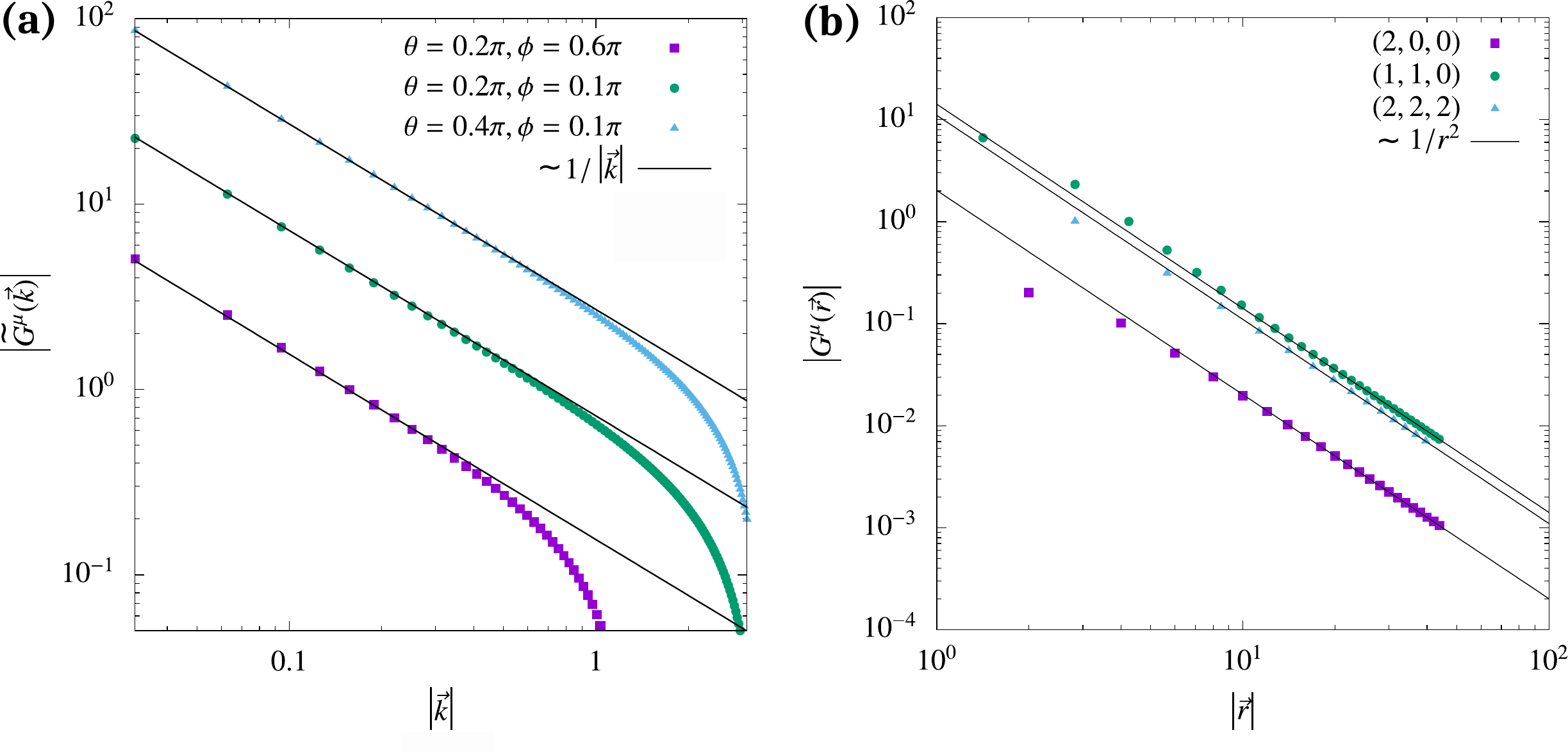}
\caption{\textbf{(a)} The radial dependence of the Green's functions in $\vec{k}$-space. The three different curves correspond to three arbitrary directions in $\vec{k}$-space along which the absolute value of the Green's function in Fourier space is evaluated and plotted in a $\log-\log$ scale in order to highlight the small $|\vec{k}|$ behaviour. \textbf{(b)} The radial dependence of the Green's functions in real space along three arbitrary lattice directions.}
\label{fig_green}
\end{figure}

For an FCC lattice, the Green's functions in $\vec{k}$-space defined in equation~\eqref{greenk} are evaluated at different values of $\vec{k}$ by performing the matrix inversions given in equation~\eqref{Greenk matrix}.
We notice that in Fourier space the Green's functions have the following behaviour in the limit of small $|\vec{k}|$
\begin{equation}
    \tilde{G}^{\mu}(\vec{k}) \sim\frac{f^{\mu}(\theta_k,\phi_k)}{g(\theta_k,\phi_k)}\frac{1}{|\vec{k}|}.
\end{equation}
where the functions $f^{\mu}$ and $g$ depends only on the angular coordinates and are given in \ref{appendix:A}. The $1/|\vec{k}|$ dependence of the Green's function in Fourier space is displayed in ~\fref{fig_green}.  Finally, an inverse Fourier transform of the above equation yields the asymptotic behaviour of the Green's function in real space at large distances as
\begin{equation}
    G^{\mu }(\vec{r})\sim\frac{1}{(2\pi)^3}\int_{0}^{2\pi}\int_{0}^{\pi}d\phi_kd\theta_k \frac{f^{\mu}(\theta_k,\phi_k)}{g(\theta_k,\phi_k)}
   \left[ \int_{\lambda}^{\pi}dk k^2 \frac{e^{-i \vec{k}.\vec{ r}}}{k}\right].
   \label{eq_greenr}
\end{equation}

While performing the integration along the radial coordinate in Fourier space $(|\vec{k}|)$ we exclude the singular point $|\vec{k}|=0$. We choose a lower cutoff of the integration, $\lambda$ corresponding to the first non-zero wavevector. This depends on the linear system size ($L$) as $\lambda \to \frac{2\pi}{L} \Lambda$ where $\Lambda$ is the lattice spacing in the underlying crystal structure. Therefore, for an infinitely large system ($L\to \infty$) where $\lambda \to 0$, the radial integration can be written as,
 \begin{equation}
 \begin{aligned}
      \lim_{\lambda\to 0}\int_{\lambda}^{\pi}dk k^2 \frac{e^{-i \vec{k}.\vec{ r}}}{k}=&\frac{-1}{i A}\frac{d}{dr}\left[\lim_{\lambda\to 0}\int_{\lambda}^{\pi} dk \exp(-i k r A(\theta,\phi,\theta_k,\phi_k))\right]\\
      =&\frac{d}{dr}\left[\frac{ \exp(-i \pi r A(\theta,\phi,\theta_k,\phi_k))-1}{  r A^2(\theta,\phi,\theta_k,\phi_k)}\right]\\
      \sim & \frac{1}{r^2},
 \end{aligned}
 \label{eq_radial_int}
 \end{equation}
where  $A(\theta,\phi,\theta_k,\phi_k)=\sin\phi_k \sin\phi \cos(\theta-\theta_k)+\cos\phi_k \cos\phi$. From equations~\eqref{eq_greenr} and ~\eqref{eq_radial_int} we see that at large distances, $G^{\mu}(\vec{r})$ scales as $1/r^2$. This dependence of the Green's function in real space is verified in ~\fref{fig_green}~(b). Similarly we can obtain the real space displacement correlations between two lattice positions $\vec{r}$ and $\vec{r}'$ situated far away from each other using equation~\eqref{eq_disp_correlation}, and by performing the radial integration we obtain,
 \begin{equation}
      \mathcal{C}_{\mu \nu}\left(\vec{r}-\vec{r}^{\prime}\right)\sim \frac{\eta^{2}}{\left|\vec{r}-\vec{r}^{\prime}\right|},
 \end{equation}

\begin{figure}[t]
  \centering
  \includegraphics[scale=0.63]{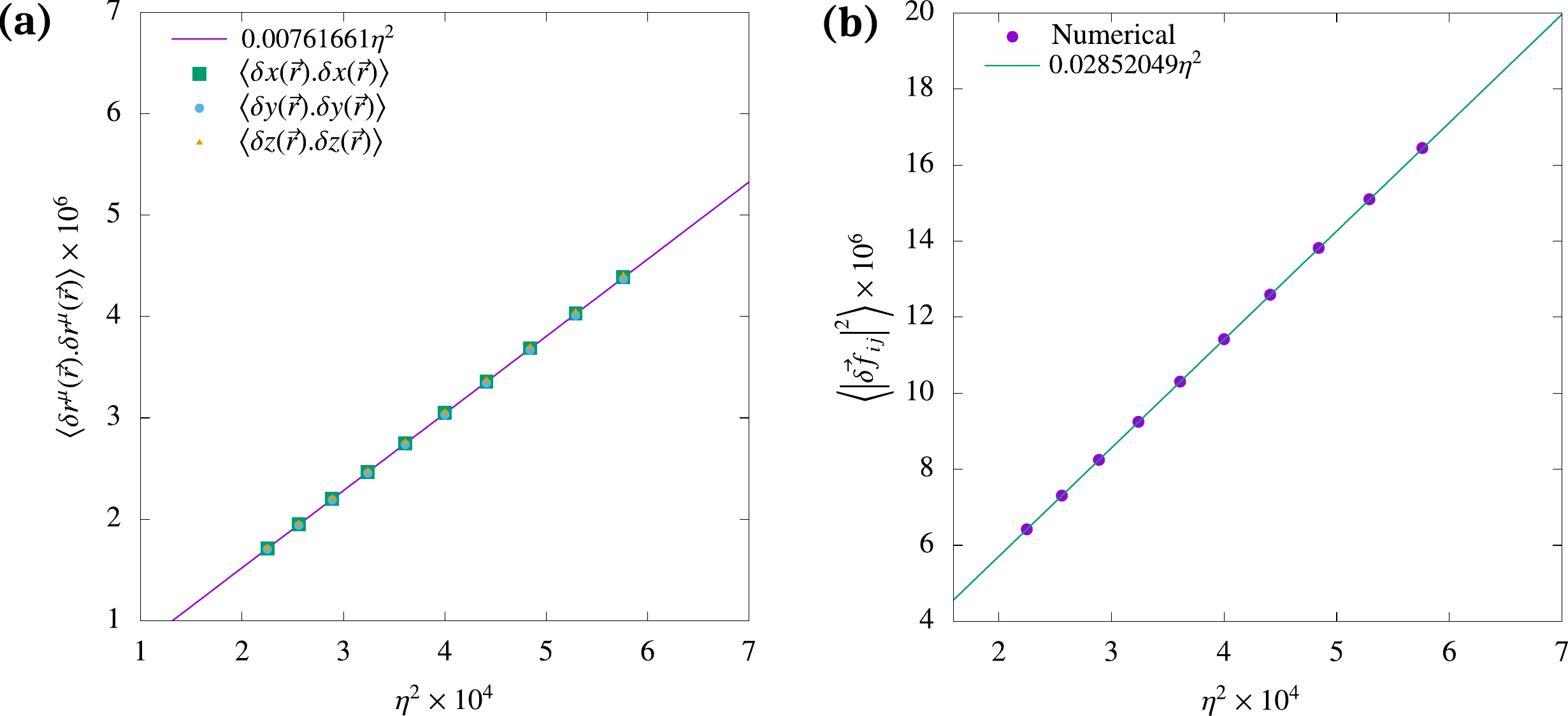}
  \caption{Comparison of numerical and analytical results, for a system of size $N = 4000$, at a packing fraction $\phi = 0.8$. \textbf{(a)} Variance in particle displacements at any lattice site, as a function of polydispersity. The theoretical predictions (solid line) are as described in equation~\eqref{eq_disp_fluctuation}. The displacement fluctuations are the same along all three lattice directions (i.e $\mu=x,y,z$) and only depends on the polydispersity ($\eta$). \textbf{(b)} Variation of average fluctuations in contact force between any two neighbouring particles given in equation~\eqref{eqs_force_fluctuation} with $\eta^2$.} 
    \label{fig_displacement_fluctuation}
\end{figure}
 
whereas the same site correlations i.e., the fluctuations in displacements at every lattice site $\vec{r}$, as a function of the polydispersity ($\eta$) is given as,
\begin{equation}
\begin{aligned}
     \mathcal{C}_{\mu \nu}\left(0\right)=&  \left\langle\delta r_{\mu}(\vec{r}) \delta r_{\nu}\left(\vec{r}\right)\right\rangle\\ =&\frac{1}{(2L)^3}\left(\frac{\eta^{2}}{48}\right) \sum_{m=0}^{2L-1} \sum_{l=0}^{2 L-1}\sum_{p=0}^{2L-1}\left(\tilde{G}^{\mu}(\vec{k}) \tilde{G}^{\nu}(-\vec{k})\right), \\
     \sim& \eta^2.
\end{aligned}
\label{eq_disp_fluctuation}
\end{equation}

The dependence of the displacement fluctuations derived above, on the polydispersity ($\eta$) is verified from numerical simulations and displayed in ~\fref{fig_displacement_fluctuation}~(a). These fluctuations are exactly described by equation~\eqref{eq_disp_fluctuation} with no fitting parameters. Similarly, the correlation between the different components of the excess forces between particles `$i$' and `$j$' can be computed by using the linearized expressions in equation~\eqref{df linear}. Upon replacing the linear displacements and particle size disorder by their Fourier transforms, we obtain
\begin{equation}
\begin{aligned}
\left\langle\delta f_{i j}^{\mu} \delta f_{i j}^{\nu}\right\rangle= \frac{1}{(2L)^3}\left(\frac{\eta^{2}}{48}\right) &\sum_{m=0}^{2L-1} \sum_{l=0}^{2 L-1}\sum_{p=0}^{2L-1}\left[\left(1-\mathcal{F}_{j}(\vec{k})\right)\sum_{\gamma}C_{i j}^{\mu \gamma} \tilde{G}^{\gamma}(\vec{k})+C_{i j}^{\mu \sigma}\left(1+\mathcal{F}_{j}(\vec{k})\right)\right]\times \\& \left[\left(1-\mathcal{F}_{j}(\vec{k})^{-1}\right) \sum_{\gamma}C_{i j}^{\nu \gamma} \tilde{G}^{\gamma}(-\vec{k})+C_{i j}^{\nu \sigma}\left(1+\mathcal{F}_{j}(\vec{k})^{-1}\right)\right].
\end{aligned}
\label{eqs_force_fluctuation}
\end{equation}

For $\mu \neq \nu $, the above correlation vanishes. The force-magnitude fluctuation between particles `$i$' and `$j$' which are neighbours in the underlying FCC arrangement is the sum of the fluctuations in forces along all the coordinate axes. This can be represented as
\begin{equation}
 \left\langle \left|\delta \vec{f}_{i j}\right|^2\right\rangle = \sum_{\mu} \left\langle\delta f_{i j}^{\mu} \delta f_{i j}^{\mu}\right\rangle.
 \label{eqs_force_parallel}
\end{equation}

From  equations~\eqref{eqs_force_fluctuation} and~\eqref{eqs_force_parallel} we obtain $ \left\langle \left|\delta \vec{f}_{i j}\right|^2\right\rangle \sim \eta^2$ which we further verify using numerical simulations, as shown in ~\fref{fig_displacement_fluctuation}~(b).

Next, we analyze the distribution of the force fluctuations produced by the athermal disorder. Since the component of the force parallel to the lattice direction can be written as  $f_{ij}^{\parallel} = f_0 + \delta f_{ij}^{\parallel} = f_0 + \eta\sigma_0\sum_{\vec{r}^{\prime}} C_{ij}^{||}(\vec{r}_i-\vec{r}^{\prime})\zeta\left(\vec{r}^{\prime}\right)$, it is also a random variable governed by a generalized Irwin-Hall distribution. Since the exact form of the distribution in equation~\eqref{eq_irwin1} involves computing increasingly large summations with increasing system sizes, it is not feasible to obtain the exact expression for the distribution. Therefore, we calculate the distribution of $\delta f^{\parallel}_{ij}$ as an inverse Fourier transform of the generalized Irwin-Hall distribution in Fourier space which has a compact expression given as,
\begin{equation}
\begin{aligned}
&\tilde{P}_L^{l}(k)=\exp\left[- \sum_{n=1}^{l}\alpha_n \left(\sum_{m=1}^{N}(\eta \sigma_0 C_{ij}^{\parallel}(\vec{r}_i-\vec{r}_m))^{2n}\right) k^{2n} \right],
\end{aligned}
\label{eq_fourier_distribution}
\end{equation}
where the limit of the summation `$l$' denotes the number of terms in the truncated series representation. We may then obtain the distribution of $\delta f^{\parallel}$ as,
\begin{equation}
\begin{aligned}
&P_{L}^{l}(\delta f^{\parallel})=\frac{1}{2\pi} \int_{-\infty}^{\infty} dk e^{ik \delta f^{\parallel}} \tilde{P}^{l}(k),
\\
&P_{L}(\delta f^{\parallel})=\lim_{l\to \infty}P_{L}^{l}(\delta f^{\parallel}).
\end{aligned}
\end{equation}

\begin{figure}[t]
\centering
\includegraphics[scale=0.23]{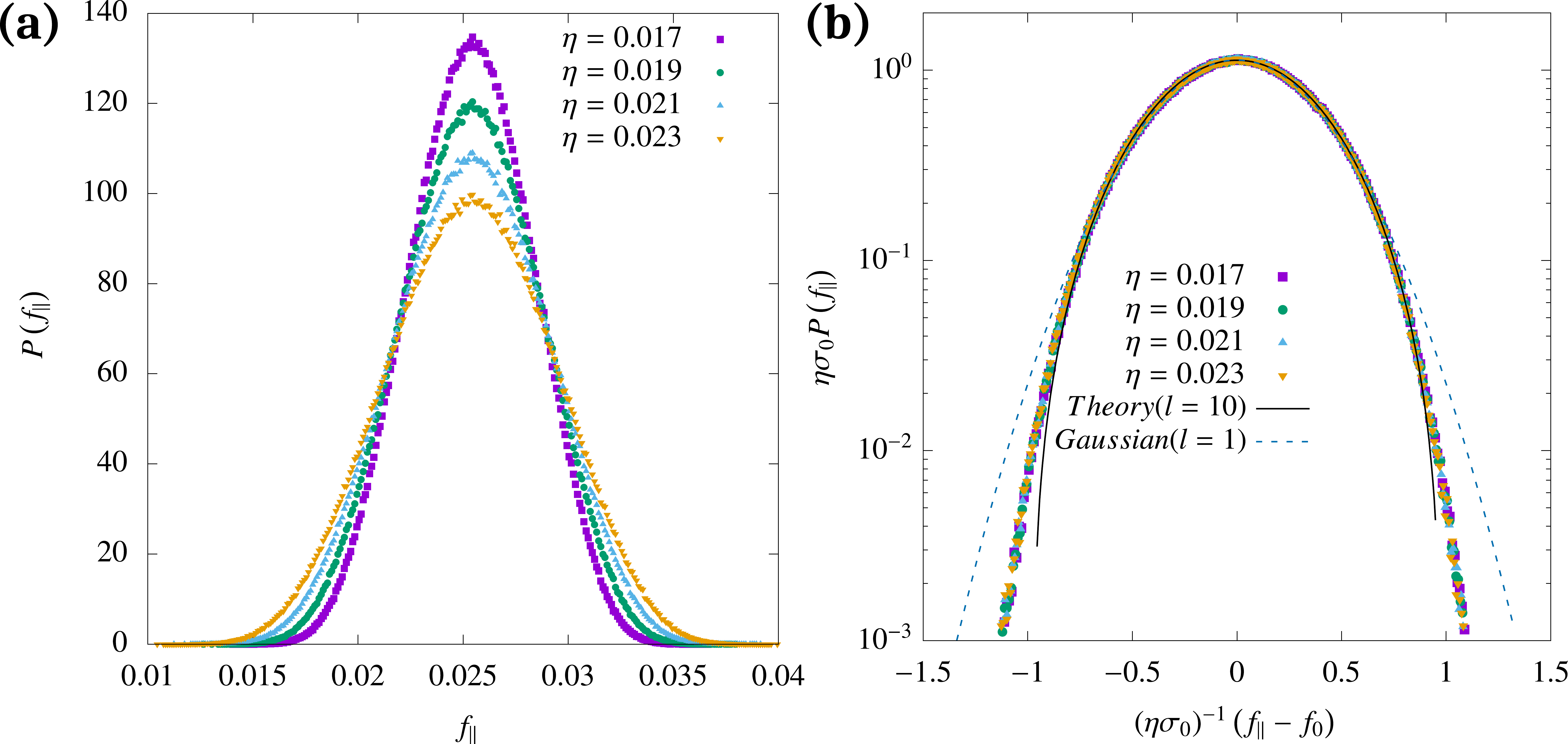}
\caption{\textbf{(a)} Probability distributions of the parallel component of forces between neighbouring particles in energy minimized configurations plotted for different polydispersities ($\eta$). \textbf{(b)} Scaled distributions with a scaling variable $(\eta \sigma_0)^{-1}$. Analytic predictions (solid line) using the Irwin-Hall distribution ($l=10$) (without any fitting parameters), deviating significantly from Gaussian ($l=1$) behaviour, at the tails of the distributions.}
\label{fig_fparallel}
\end{figure}

Finally, we can obtain the probability distribution function (PDF) of $f^{\parallel}$ by shifting the above distribution by $f_0$. The accuracy of the distributions may be improved by retaining more terms in equation~\eqref{eq_fourier_distribution}. Here, we choose $l=10$ in the expression and compare it against results from simulations, as shown in ~\fref{fig_fparallel}(b).  Crucially, keeping the first term in equation~\eqref{eq_fourier_distribution} yields a Gaussian distribution, as observed in two dimensions in reference~\cite{acharya2020athermal}. However, as the number of terms is increased the tails of the distribution differ significantly from a Gaussian, as is borne out from our numerical simulations.

We now consider the perpendicular component of the interparticle forces. For convenience, we consider a bond in the $xz$-plane along a fixed direction i.e. $(1,0,1)$ in the crystalline state. If the system is rotated about the y-axis by an angle $\theta=\pi/4$ in the counterclockwize direction, then all the bonds along ($1,0,1$) will be aligned with the $z$-axis. Therefore in the rotated frame, the $z$-axis is aligned along one of the lattice directions of the crystalline state. Upon introducing disorder, this bond will contain force components both along and perpendicular to the $z$-axis. Therefore the force components in the rotated frame can be written as,
\begin{equation}
\left(\begin{array}{c}
f^{'x}_{ij} \\
f^{'y}_{ij}\\
f^{'z}_{ij}
\end{array}\right)=
\left(\begin{array}{ccc}
\cos(\pi/4) & 0 & -\sin(\pi/4)\\
0 & 1 & 0 \\
\sin(\pi/4) & 0 & \cos(\pi/4)
\end{array}\right)\left(\begin{array}{c}
f^{x}_{ij} \\
f^{y}_{ij}\\
f^{z}_{ij}
\end{array}\right)=\left(\begin{array}{c}
(f^{x}_{ij}-f^{z}_{ij})/\sqrt{2}\\
f^{y}_{ij}\\
(f^{x}_{ij}+f^{z}_{ij})/\sqrt{2}
\end{array}\right).
\end{equation}

\begin{figure}[t!]
\centering
\includegraphics[scale=0.125]{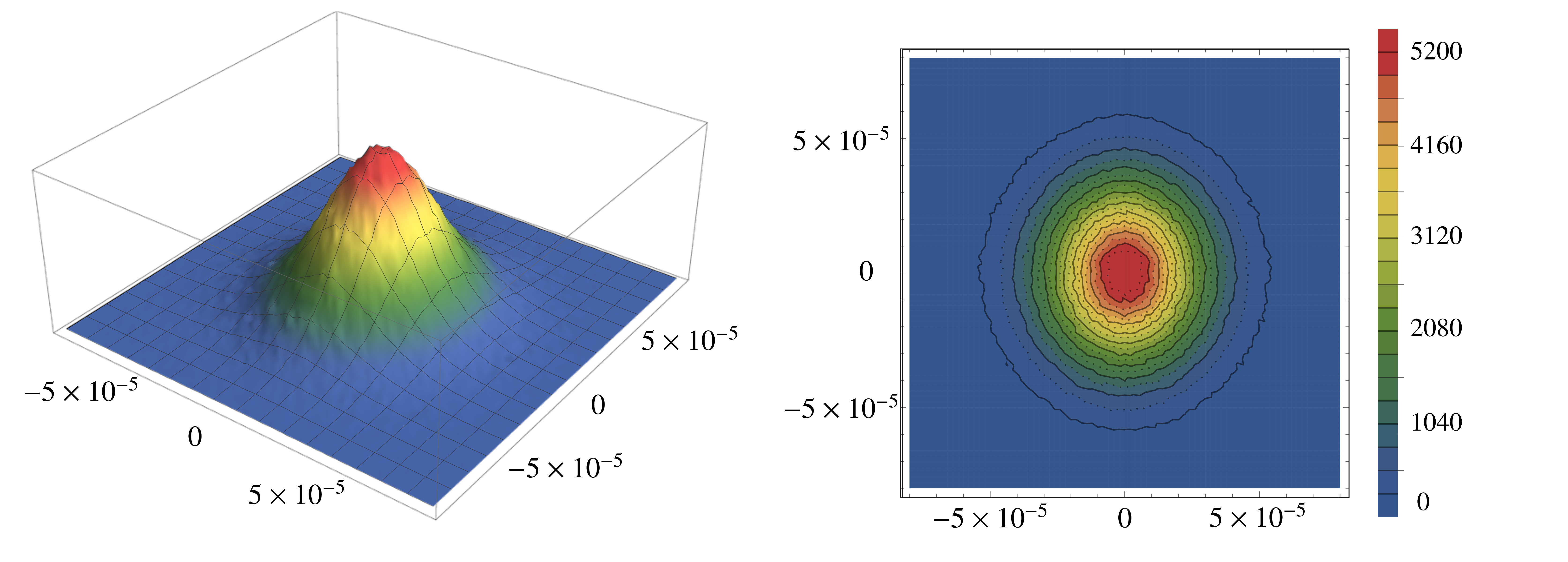}
\caption{Probability distribution of the force component between neighbouring particles in energy minimized configurations that lies on the plane perpendicular to the lattice directions. These results correspond to a system with polydispersity $\eta =0.015$.}
\label{fig_fperpendicular}
\end{figure}

From the above equation we see that the $f_{ij}^{'z}= \vec{f}_{ij}\cdot \vec{r}_{ij}= (f^{x}_{ij}+f^{z}_{ij})/\sqrt{2}$, represents the component of the forces parallel to the initial bond direction, whereas $(f^{'x}_{ij}, f^{'y}_{ij})$ are the perpendicular components. Also note that the widths of $f^{'x}_{ij}$ and $f^{'y}_{ij}$ display an anisotropy, as may be seen in ~\fref{fig_fperpendicular}. This asymmetry captures the different environments in the two perpendicular directions in an FCC lattice. Lastly, comparing the fluctuations in parallel and perpendicular components of interparticle forces, we observe that the perpendicular component is far more constrained (width$\sim 10^{-5}$) as compared to its parallel component (width $\sim 10^{-2}$), as has also been observed in 2D, in reference~\cite{acharya2020athermal}. This is also clear from the large difference in the spread of both the distributions in figures~\ref{fig_fparallel} and ~\ref{fig_fperpendicular}.

\section{Conclusions and Discussion}

In this paper, we have studied the effects of introducing disorder into a three dimensional athermal crystalline system, both analytically as well as numerically. The disorder, which forms a tunable parameter, is manifest \textit{via} a polydispersity $(\eta)$ in particle sizes. In the regime of small disorder, we have performed a perturbative analysis to derive exact results corresponding to the displacement of particles as well as the excess interparticle forces between particles. Specifically, we have shown that $\delta \vec{r} \propto \eta$ and  $\delta \vec{f}_{ij} \propto \eta$. We have also derived exact expressions for the fluctuations of these quantities as a function of the disorder, $\left\langle\delta r_{\mu}(\vec{r}) \delta r_{\nu}\left(\vec{r}\right)\right\rangle \propto \eta^2$ and $\left\langle\delta f_{ij}^{\mu}(\vec{r}) \delta f_{ij}^{\nu}\left(\vec{r}\right)\right\rangle \propto \eta^2$. We have numerically verified our analytic results for an FCC lattice. In conjunction with recent studies of triangular lattices in two dimensions~\cite{acharya2020athermal,acharya2021disorder,das2021long}, our work highlights the generality of the method, which is valid across dimensions. We have further explored the displacement correlation functions, $\left\langle\delta r_{\mu}(\vec{r}_0) \delta r_{\nu}\left(\vec{r}_0+\vec{\Delta}\right)\right\rangle$, which varies in three dimensions, as $\sim 1/\Delta$, where $\Delta$ is the magnitude of the separation between particles. This is in contrast with the correlation in two dimensions, where it varies logarithmically with separation $( \sim -\log(\Delta))$~\cite{acharya2021disorder} for small $\Delta$.
Finally, we have also derived exact analytical forms of the distributions of the displacement and force fluctuations, which show clear deviations from Gaussian behaviour, significant in the tails of distributions. It would be interesting to extend these results to other stable lattices in three and higher dimensions, possibly with the addition of next-nearest neighbour interactions.

\ack
Useful discussions with Vishnu V.~Krishnan, Kabir Ramola, Stephy Jose, Pappu Acharya, Debankur Das, Soham Mukhopadhyay and Surajit Chakraborty are gratefully acknowledged. This project was funded by intramural funds at TIFR Hyderabad from the Department of Atomic Energy (DAE).

\appendix

\section{Angular dependence of the Green's function in the Fourier space
}\label{appendix:A}
\setcounter{figure}{0}

In this appendix, we derive the expression for the Green's function for displacement fields in a system of over compressed soft particles. The particles are arranged in FCC lattice and interact with a potential given in equation~\eqref{potential} with $\alpha =2$ and $K_{ij}=1/2$. In Fourier space the Green's function for the displacement field can be represented as, 
\begin{equation}
    \tilde{G}^{\mu}(\vec{k})=\sum_{\nu}\tilde{\mathcal{G}}^{\mu \nu}(\vec{k}) D^{\nu}(\vec{k})=\sum_{\nu}(A^{-1})^{\mu \nu} D^{\nu},
    \label{Greenk_appendix}
\end{equation}

where the expressions for both $\tilde{\mathcal{G}}^{\mu \nu}$ and $D^{\nu}$ are provided in equation~\eqref{Amat}. Performing the inversion of the matrix $\textbf{A}$ yields the matrix $\tilde{\mathcal{G}}$. The elements of both $\textbf{A}$ and $\textbf{D}$ contain terms involving $C_{ij}^{\mu \nu}$, which depends on the underlying lattice and for FCC lattice, these elements can be expressed as
\begin{equation}
    \begin{aligned}
     A^{\mu\nu}=&\frac{\alpha(1-R_0)^{\alpha-2}(1+R_0(\alpha-2))\sin(k^{\mu})\sin(k^{\nu})}{R_0}\\
    A^{\mu\mu}=&\frac{\alpha(1-R_0)^{\alpha-2}}{R_0}\left[2(1-R_0)(1-\cos(k^{\nu})\cos(k^{\gamma}))+\right.\\
    &\hspace{2cm}\left.(1-\alpha R_0)(2-\cos(k^{\mu})\cos(k^{\nu})-\cos(k^{\mu})\cos(k^{\gamma}))\right]\\
    D^{\mu}=&-i\sqrt{2}\alpha(1-R_0)^{\alpha-2}(1-\alpha R_0)\sin(k^{\mu})(\cos(k^{\nu})+\cos(k^{\gamma})).
    \end{aligned}
    \label{AandD_element}
\end{equation}
 
In equation~\eqref{AandD_element}, $\mu,\nu, \gamma \in \{x,y,z\}$ and $\mu\neq \nu \neq \gamma$. Since for small $|\vec{k}|$, $A^{\mu \nu} \propto |\vec{k}|^2 $ and $D^{\mu} \propto |\vec{k}|$, we can obtain the small $|\vec{k}|$ behaviour of the Green's function $\tilde{G}^{\mu}(\vec{k})$ as  $\tilde{G}^{\mu}(\vec{k}) \propto 1/|\vec{k}|$. Next, we obtain the exact expression for the Green's function in the Fourier space by performing the summation in equation~\eqref{Greenk_appendix}, which
has the following small $|\vec{k}|$
behaviour
\begin{equation}
    \tilde{G}^{\mu}(\vec{k}) \sim\frac{f^{\mu}(\theta_k,\phi_k)}{g(\theta_k,\phi_k)}\frac{1}{|\vec{k}|},
\end{equation}
where the common denominator for all the three Cartesian directions corresponding to three different components of displacement fields in the angular function has the form
\begin{equation}
\begin{aligned}
    &g(\theta_k,\phi_k)=-8 \cos(\theta_k)^4 (-3 + 4 R_0)\{11-(44-32R_0)R_0+(1+4R_0)\cos(2\theta_k)\}  -\\
    &(-5+8R_0+\cos(2\theta_k))\left[\{(45-4R_0(5+3R_0)+2R_0(-70+70R_0)+3\cos(4\phi_k)\}\sin(\theta_k)^4\right.\\
   & \left.+2(-3+4R_0)(-5+8 R_0)\sin(2\theta_k)^2  \right]
    +16\cos(\theta_k)^2\sin(\theta_k)^4(-13+16R_0+\cos(4\phi_k)).
\end{aligned}
\end{equation}
Now, the direction-dependent terms in the numerator can be written as

\begin{equation}
\begin{aligned}
    f^x(\theta_k,\phi_k)=&-4\sqrt{2}\cos(\phi_k)\sin(\theta_k) R_0(-1+2R_0)(7-8R_0+\cos(2\theta_k))\\
    &(26-32R_0-2\cos(2\theta_k)-4\cos(2\phi_k)\sin(\theta_k)^2),\\
    f^y(\theta_k,\phi_k)=&-4\sqrt{2}\sin(\phi_k)\sin(\theta_k) R_0(-1+2R_0)(7-8R_0+\cos(2\theta_k))\\
     &(26-32R_0-2\cos(2\theta_k)-4\cos(2\phi_k)\sin(\theta_k)^2),\\
     f^z(\theta_k,\phi_k)=&-\sqrt{2}\cos(\theta_k) R_0(-1+2R_0)\left[ (-27+32R_0)(-25+32R_0)+\right.\\
     &\left. 4(-25+32R_0)\cos(4\theta_k)-8\cos(4\phi_k)\sin(\theta_k)^4  \right].\\
\end{aligned}
\end{equation}

\begin{figure}[t]
\hspace*{-0.3cm}
\includegraphics[scale=0.192]{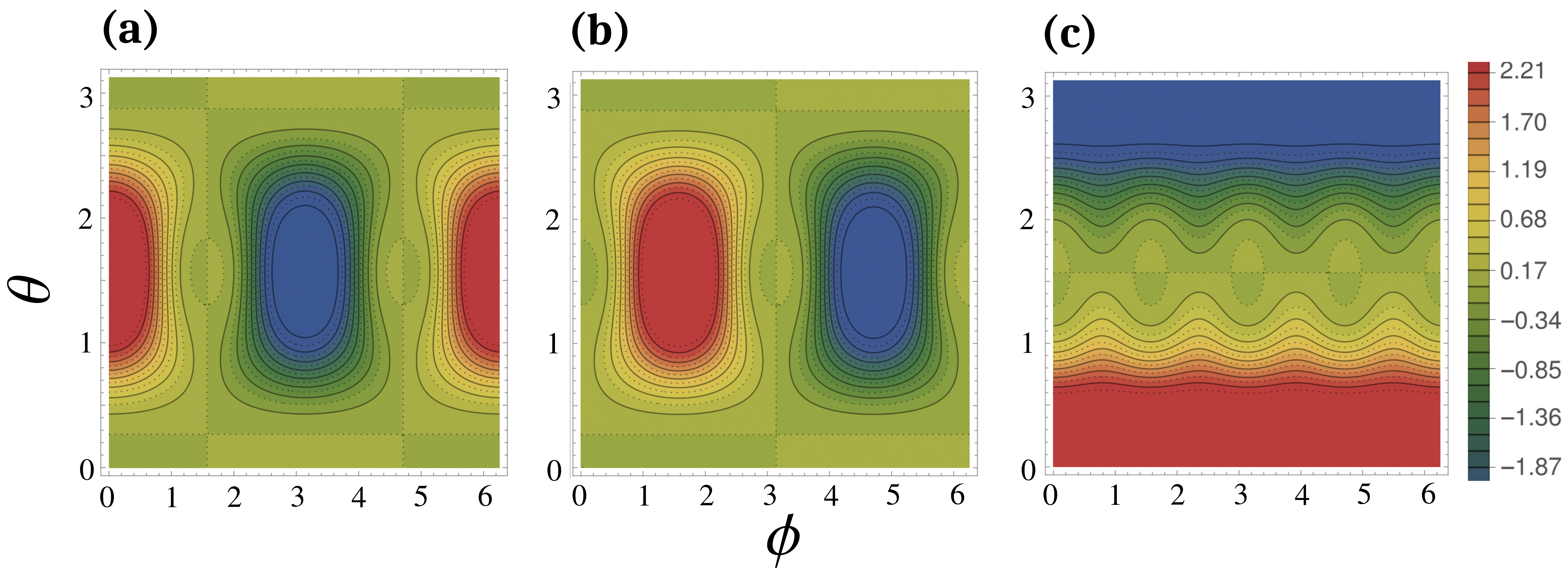}
\caption{Density contour plot of the angular dependence of the Green's functions in Fourier space i.e. $\frac{f^{\mu}(\theta,\phi)}{g(\theta,\phi)}$ in equation~\eqref{Greenk_appendix}. Here the $x$-axis represents the azimuthal angle $\phi$ and the $y$ axis represents the polar angle $\theta$. }
\label{fig_angular_green}
\end{figure}

\section{Derivation of the probability distribution of a sum of uniformly distributed random numbers with multiplicative coefficients}
\label{appendix:B}

Let $x_i$ for $i=1,2,\cdots, N$ be independent random variables uniformly distributed in the interval $[a_i,b_i]$, and let $z=\sum_{i=1}^{N}G_ix_i$. Then we can make a change of variables as,
 \begin{equation}
     X_i=\frac{x_i-a_i}{b_i-a_i}.
     \label{X_i appendix}
 \end{equation}
Here $X_i$ are uniform random numbers in the interval $[0,1]$. Multiplying equation~\eqref{X_i appendix} by $G_i$ and summing over $i$ yields 
 \begin{equation}
     \sum_{i=1}^{N}G_i X_i=\sum_{i=1}^{N}\frac{G_i (x_i-a_i)}{b_i-a_i}=t.
     \label{t Appendix}
 \end{equation}
 
   For $m \in\{0,1,2, \ldots, N-1\}$ and $t \in \left[\sum_{i=1}^{m}G_i , \sum_{i=1}^{m+1}G_i \right)$, we define the sets
\begin{equation}
\begin{array}{l}
A_{N}(t)=(\left(X_{1}, X_{2}, \ldots, X_{N}\right): X_{i} \geq 0, \text {      for       } i \in\{1,2, \ldots, N\}, \sum_{i=1}^{N} G_i X_{i} \leq t ), \\
B_{j}(t)=\left\{\left(X_{1}, X_{2}, \ldots, X_{N}\right) \in A_{N}(t): X_{j}>1\right\}, \\
C_{N}=\left\{\left(X_{1}, X_{2}, \ldots, X_{N}\right): 0 \leq X_{i} \leq 1\right\},
\end{array}
\label{hypervolumes_def}
\end{equation}
where $C_N$ is the $N$-dimensional unit cube. The complement set of $C_{N}$ is denoted as $C_{N}^{\prime}$. The hypervolume of the $N$-dimensional solid $A_{N}(t)$ given in equation~\eqref{hypervolumes_def} can be written as
\begin{equation}
\operatorname{Vol}\left(A_{N}(t)\right)=\iiint\limits_{\sum_{i=1}^{N} G_i X_{i} \leq t }dX_1 dX_2 \ldots dX_N=\frac{t^{N}}{N !\prod_{i=1}^{N}G_i }.
\end{equation}
\begin{figure}[t!]
\centering
\includegraphics[width=0.60\linewidth]{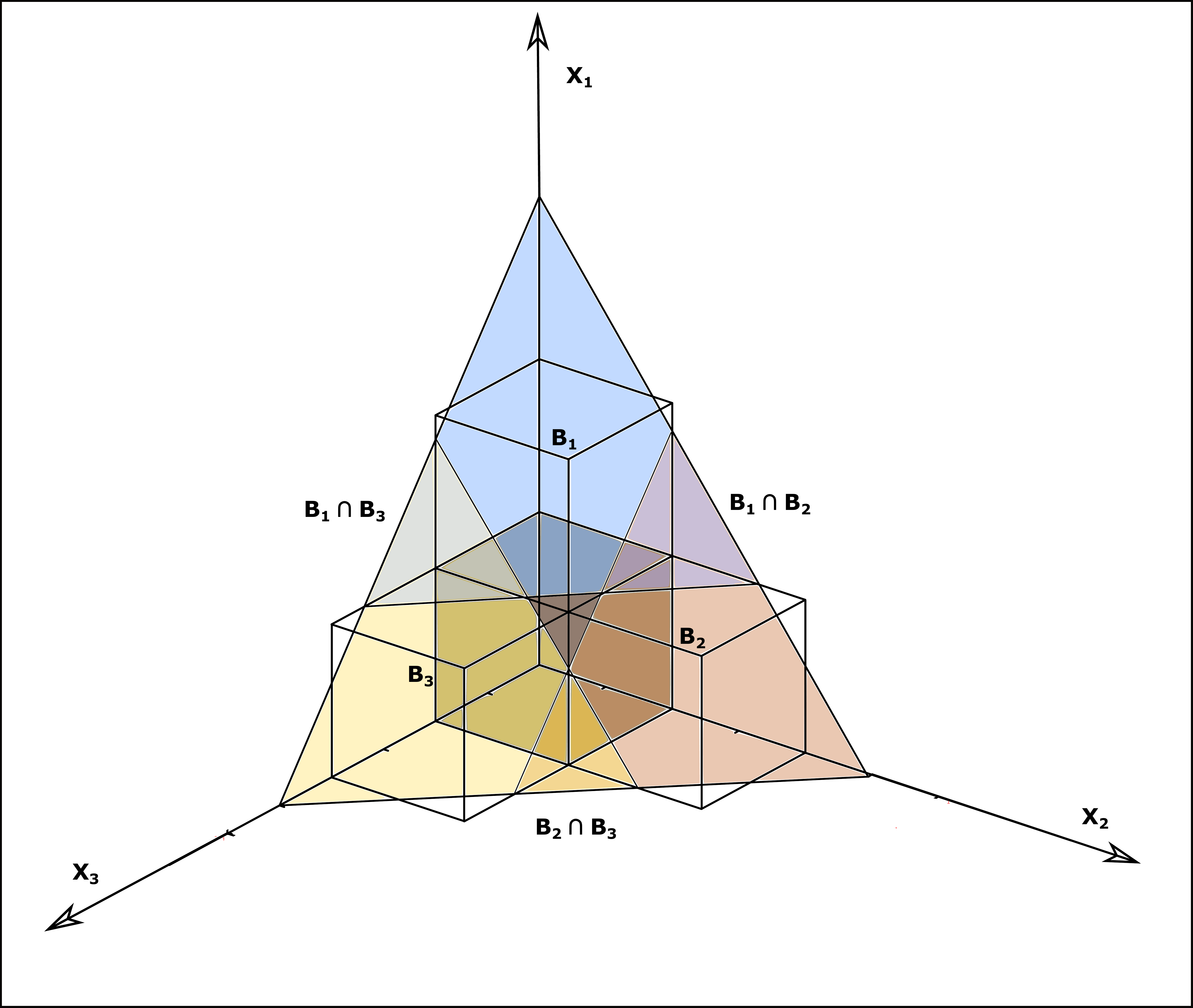}
\caption{A schematic diagram of different subregions of a 3D simplex.}
\label{Schematic_simplex}
\end{figure}

 This $N$-dimensional solid is a standard orthogonal simplex from the corner of an $N$-dimensional hypercube. We next consider a simplex defined by a region  where $X_{j}>1 $ for $j\in\{1,2, \ldots, k\}$ and  $0 \leq X_{j} \leq 1 $ for $j\in\{k+1,k+2, \ldots, N\}$. These regions for the case of 3 variables i.e ($X_1,X_2,X_3$) are displayed in ~\fref{Schematic_simplex}. To find the length of the sides of the simplex in this region represented as $\bigcap_{j=1}^{k} B_{j}(t)$, we solve the linear equations in the plane 
 \begin{equation}
\begin{array}{l}
 X_n=\frac{(t-\sum_{i=1,i\neq n}^{k}G_i)}{G_n}  \text {,     for     } n \in \{1,2, \ldots, k\},\\X_m=\frac{(t-\sum_{i=1}^{k}G_i)}{G_m}  \text {,     for     } m \in \{k+1,k+2, \ldots, N\}  .
\end{array}
\end{equation}
 The length of the sides of the simplex in the region defined above is thus given as,
\begin{equation}
\begin{array}{l}
 L_n=X_n-1=\frac{(t-\sum_{i=1}^{k}G_i)}{G_n} \text {,     for     } n \in \{1,2, \ldots, k\},\\ L_m=X_m=\frac{(t-\sum_{i=1}^{k}G_i)}{G_m} \text {,     for     } m \in \{k+1,k+2, \ldots, N\} .
\end{array}
\end{equation}
 For $k \in\{1,2, \ldots, m\}$,  the hypervolume of $\bigcap_{j=1}^{k} B_{j}(t)$ is
\begin{equation}
\operatorname{Vol}\left(\bigcap_{j=1}^{k} B_{j}(t)\right)=\frac{\prod_{n=1}^{N}L_n}{N !}=\frac{(t-\sum_{n=1}^{k}G_n)^N}{N !\prod_{n=1}^{N}G_n},
\end{equation}
and for $k \in\{m+1, m+2, \ldots, n\}$,
\begin{equation}
\operatorname{Vol}\left(\bigcap_{j=1}^{k} B_{j}(t)\right)=0.
\end{equation}
We define a function,
\begin{equation}
\begin{aligned}
f(x) &=x ,\text { for } x > 0, \\
&=0,\text { for } x \leq 0.
\end{aligned}
\end{equation}
Now, we rewrite the hypervolumes in each subregion using this function as
\begin{equation}
\operatorname{Vol}\left(\bigcap_{j=1}^{k} B_{j}(t)\right)=\frac{\left(f\left(t-\sum_{n=1}^{k}G_n\right)\right)^N}{N !\prod_{n=1}^{N}G_n}.
\end{equation}
Using the inclusion-exclusion principle, we can find the cumulative distributive function (CDF) of `$t$' satisfying the condition in equation~\eqref{t Appendix} as
\begin{equation}
\begin{aligned}
    F(t)&=P(T \leq t)=\operatorname{Vol}\left(A_{N}(t) \cap C_{N}\right)\\
   &=\operatorname{Vol}\left(A_{N}(t)\right)-\operatorname{Vol}\left(A_{N}(t) \cap C_{N}^{\prime}\right)=\operatorname{Vol}\left(A_{N}(t)\right)-\operatorname{Vol}\left(\bigcup_{j=1}^{N} B_{j}(t)\right)\\
    &=\frac{t^{N}}{N !\prod_{i=1}^{N}G_i }-\sum_{k=1}^{N}(-1)^{k-1}\left[
 \sum_{1 \leq j_{1}<j_{2}<\cdots<j_{k} \leq N} \operatorname{Vol}\left(B_{j_{1}}(t) \cap B_{j_{2}}(t) \cap \cdots \cap B_{j_{k}}(t)\right)\right]\\
 &=\frac{t^{N}}{N !\prod_{i=1}^{N}G_i }-\sum_{k=1}^{N}(-1)^{k-1}\left[
 \sum_{1 \leq j_{1}<j_{2}<\cdots<j_{k} \leq N} \frac{\left(f\left(t-\sum_{n=1}^{k}G_{j_n}\right)\right)^N}{N !\prod_{n=1}^{N}G_{j_n}}\right]\\
 &=\frac{1}{N !\prod_{i=1}^{N}G_i }\left[t^{N}-\sum_{k=1}^{N}(-1)^{k-1} g_k(t)
 \right],
\end{aligned}
\label{F(t) Appendix}
\end{equation}
where 
\begin{equation}
    g_k(t)=\sum_{1 \leq j_{1}<j_{2}<\cdots<j_{k} \leq N} \left(f\left(t-\sum_{n=1}^{k}G_{j_n}\right)\right)^N.
\end{equation}

Now we derive the probability distribution function (PDF) by taking a derivative of equation~\eqref{F(t) Appendix} with respect to $t$ and thus arrive at
\begin{equation}
    P(t)=\frac{1}{(N-1) !\prod_{i=1}^{N}G_i }\left[t^{N-1}-\sum_{k=1}^{N}\frac{(-1)^{k-1}}{N}
 \frac{d g_k(t)}{dt}\right].
\end{equation}

The above distribution is a generalisation  of the Irwin-Hall distribution. If each random variable $x_i$ is distributed in the same domain $[a,b]$ then  $t=\frac{z}{b-a}-\frac{ a}{b-a}\sum_{i=1}^{N}G_i$. Now the CDF for the variable $z$ that satisfies the condition $z=\sum_{i=1}^{N}G_ix_i$ has the form

\begin{equation}
\begin{aligned}
     &F(z)=\frac{1}{N!\prod_{i=1}^{N}G_i }\left[\left(\frac{ z-a\sum_{i=1}^{N}G_i}{b-a}\right)^{N-1}-\sum_{k=1}^{N}(-1)^{k-1}
g_k\left(\frac{ z-a\sum_{i=1}^{N}G_i}{b-a}\right)\right].
\end{aligned}
\end{equation}

We have therefore succeeded in performing an exact geometrical derivation of the generalized Irwin-Hall distribution. Since this expression is quite complicated, it may be difficult to use this in situations where $N$ is very large. Therefore we can use a Fourier transfer technique derived in reference~\cite{maharana2022first} to approximate the above PDF into the following compact form  

\begin{equation}
    P(z)= \lim_{l\to \infty}P_N^{l}(z)=\frac{1}{2\pi} \int_{-\infty}^{\infty} ds e^{ikz} \tilde{P}_{N}^{l}(k),
\end{equation}
where the partial Fourier transform of the distribution is given by
\begin{equation}
     \tilde{P}_{N}^{l}(k)=\exp\left[- \sum_{n=1}^{l}\alpha_n \left(\sum_{i=1}^{N}G_i^{2n}\right) k^{2n} \right].
\end{equation}
As we consider more number of terms ($l$) in the summation we get more accurate results for this distribution, as described in the main text.

\section*{References}
\bibliography{bibliography.bib}{}
\bibliographystyle{iopart-num}
\end{document}